\documentclass[fleqn,usenatbib]{mnras}
\usepackage{newtxtext,newtxmath}

\usepackage{lscape}
\usepackage[T1]{fontenc}
\usepackage{ae,aecompl}
\usepackage{tabularx}
\usepackage{epsfig}
\usepackage{graphicx}
\usepackage{float}
\usepackage{amsmath}
\usepackage{mathrsfs}
\usepackage{mathtools}
\usepackage{afterpage}
\usepackage{epsf}
\usepackage[normalem]{ulem}

\usepackage{graphicx}
\usepackage[colorinlistoftodos]{todonotes}

\title[Diamagnetic Accretion]{The Effect of a Magnetic Field on the Dynamics of Debris Discs Around White Dwarfs}

\author[Hogg\ Cutter \& Wynn]{M~A~Hogg \href{https://orcid.org/0000-0001-6407-1615}{\includegraphics[scale=0.05]{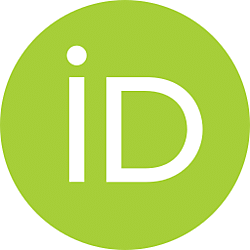}} \thanks{E-mail: mah63@le.ac.uk}$^{1}$, 
R~Cutter \href{https://orcid.org/0000-0001-8945-5551}{\includegraphics[scale=0.05]{figs/ID.png}}
\thanks{E-mail: r.cutter@warwick.ac.uk}$^{2}$, 
\& G~A~Wynn$^{1}$\vspace{0.1in}
\\
$^{1}$Theoretical Astrophysics Group, Department of Physics and Astronomy, University of Leicester, Leicester, LE1 7RH, UK
\\
$^{2}$Department of Physics, University of Warwick, Gibbet Hill Road, Coventry
CV4 7AL, UK
}

\date{Received YYY; in original form ZZZ}

\begin{document}
\maketitle

\begin{abstract}
Observational estimates of the lifetimes and inferred accretion rates from debris discs around polluted white dwarfs are often inconsistent with the predictions of models of shielded Poynting-Robertson drag on the dust particles in the discs. Moreover, many cool polluted white dwarfs do not show any observational evidence of accompanying discs. This may be explained, in part, if the debris discs had shorter lifetimes and higher accretion rates than predicted by Poynting-Robertson drag alone. We consider the role of a magnetic field on tidally disrupted diamagnetic debris and its subsequent effect on the formation, evolution, and accretion rate of a debris disc. We estimate that magnetic field strengths greater than $\sim$10kG may decrease the time needed for circularisation and the disc lifetimes by several orders of magnitude and increase the associated accretion rates by a similar factor, relative to Poynting-Robertson drag. We suggest some polluted white dwarfs may host magnetic fields below the typical detectable limit and that these fields may account for a proportion of polluted white dwarfs with missing debris discs. We also suggest that diamagnetic drag may account for the higher accretion rate estimates among polluted white dwarfs that cannot be predicted solely by Poynting-Robertson drag and find a dependence on magnetic field strength, orbital pericentre distance, and particle size on predicted disc lifetimes and accretion rates. 
\end{abstract}


\section{Introduction}
White dwarfs (WDs) are the final stage of low mass stellar evolution and are the most common outcome for stars in our Galaxy. WDs are classified by their spectra, specifically from the absorption lines from elemental abundances in their atmospheres. The DZ and DAZ classes of WDs are those observed to have metals in their atmospheres. While most WDs typically only show signatures of hydrogen or helium, around 15-35\% of the WD population show evidence of having metals in their spectra \citep{zuckerman_koester_2003, xu_jura_2012}. 
The presence of heavy metals in white dwarf atmospheres (often referred to as 'polluted' atmospheres) was originally thought to be result of accretion from the interstellar medium  or cometary impacts \citep{dupuis_fontaine_1993, alcock_fristrom_1986}, but in a seminal paper \cite{jura_2003} explained observations of G29-38 as the result of accretion from a dust disc formed by the tidal disruption of of a small, rocky asteroid. This is now the primary explanation of WD pollution.

\begin{figure}
\centering
\includegraphics[width=\columnwidth]{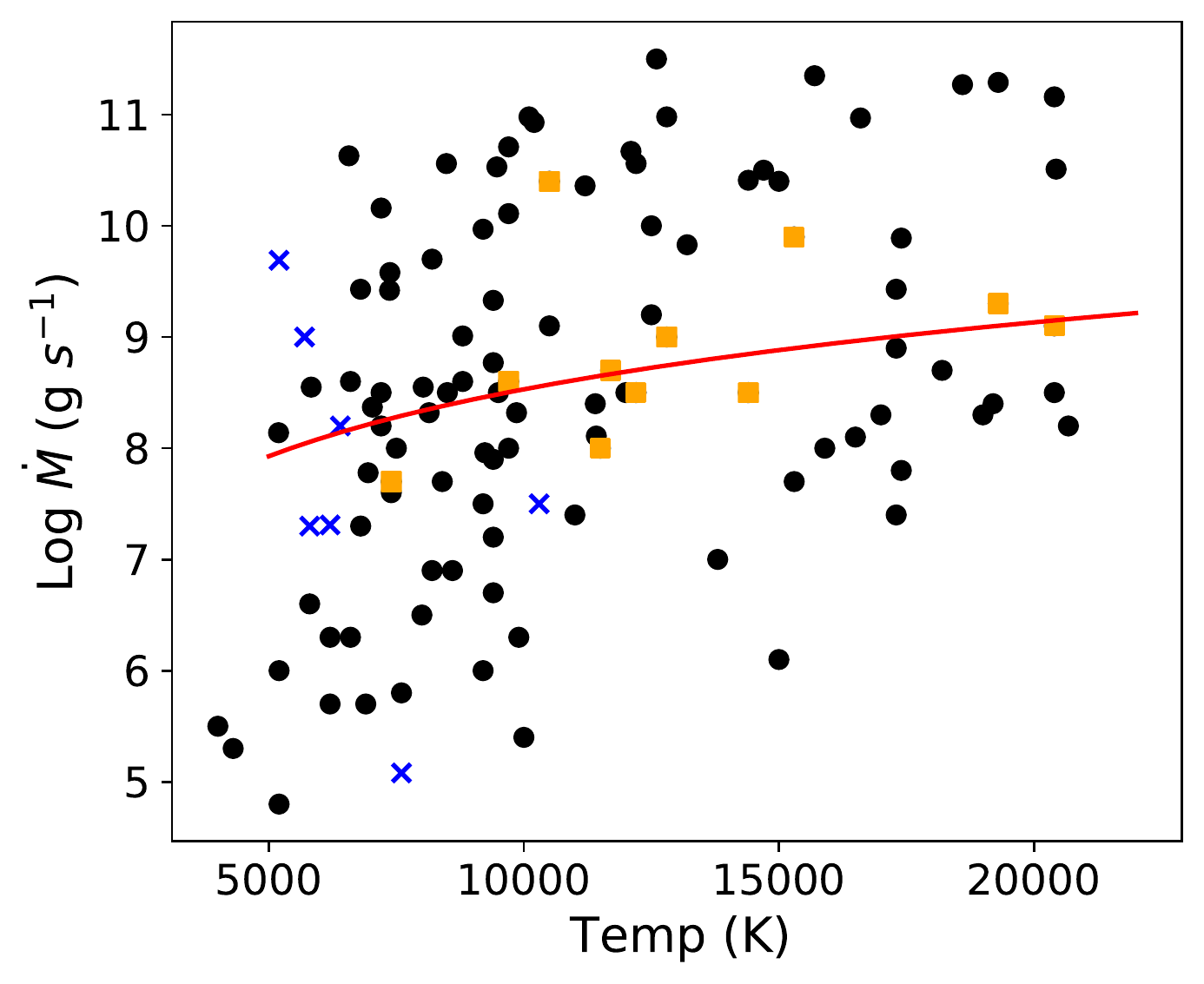}
\caption{PR drag (red line) compared to inferred DZ and DAZ accretion rates. Systems that host observable discs are marked by an orange square. The PR drag estimate is found using equations from \citet{rafikov_2011a}. The accretion rates for the individual systems are taken from \citep{friedrich_koester_2000, wolff_koester_2002, koester_wilken_2006, mullally_kilic_2007, voss_koester_2007,  dufour_bergeron_2007, desharnais_wesemael_2008, farihi_2008a, farihi_2008b, farihi_jura_2009}. Magnetic systems (blue crosses) taken  from \citep{kawka_vennes_2011, farihi_dufour_2011, zuckerman_koester_2011, KAWKA2012, kawka_vennes_2019} }
\label{'fig:PR_plot'}
\end{figure}

The accretion of an asteroid onto a WD is thought to take place following a perturbation of the orbit of the asteroid, leaving it in a highly eccentric orbit. If this orbit takes the asteroid close enough to the white dwarf it can become tidally disrupted, creating a stream of debris. \citep{debes_sigurdsson_2002}. Possible explanations for the orbital perturbation include unstable planetary systems \citep{joasil_payne_2017, mustill_villaver_2018}, eccentric planets \citep{frewen_hansen_2014} and stellar binary companions \citep{bonsor_veras_2015, veras_xu_2018}.
The processes determining the dynamics of the debris stream following the encounter between the asteroid and the WD is uncertain, including the mechanisms driving the formation and circularisation of the debris disc and the accretion of the debris by the white dwarf \citep{veras_leinhardt_2015, malamud_perets_2019a, malamud_perets_2019b}. 
Poynting-Robertson (PR) drag is thought to be an important factor in driving the accretion process. \cite{burns_lamy_1979} found the force on a particle based on incident radiation. However, when applied in the context of an "optically thin" debris disc, micron sized particles are expected to survive 10--100 years \citep{farihi_2016}. Observations of discs over decade timescales imply that debris discs persist much longer than this. To address this problem, \cite{rafikov_2011a} introduced the concept of shielding in a geometrically flat and optically thick disc, such that particles and sublimated gas on the inner rim of the disc shield the particles further out and stop them receiving the full incident starlight. This lengthens the disc lifetimes and gives accretion rates that better fit observations. This shielded regime provides a maximum limit on accretion rates as shown in Figure \ref{'fig:PR_plot'}, where the limit is shown against observational estimates of the accretion rate, highlighting that approximately 50\% of the inferred accretion rates for DZ and DAZ WDs are higher than predicted by shielded PR drag. 

Unshielded PR drag, where the disc is optically thin (in the sense that incident starlight is able to reach all dust grains within the disc), assumes the disc to be low mass, so that shielding is ineffective. Shielded PR drag, where the disc is optically thick and particles shield others from the full incident starlight, assumes higher disc masses and implies higher accretion rates \citep{rafikov_2011a, bochkarev_rafikov_2011, metzger_rafikov_2012}. Estimates based on shielded PR drag find disc lifetimes to be of the order of Myrs \citep{farihi_jura_2009}.

 The expected time taken for metals to sink out of the observable WD atmosphere (sinking timescale) varies dramatically depending on the WD, where warm hydrogen dominanted DAs have sinking timescales of order days while helium dominanted DBs and cool DAs have upper limits of order Myrs \citep{koester_2009}. This means, assuming solely shielded PR drag, there should be a large proportion of polluted WDs with discs. However, only 0.8 - 6.5\% of the polluted WD population have been found to have an observable disc \citep{debes2007, farihi_2016}.
\citet{bonsor_farihi_2017} suggest four potential ways the disc could escape detection: opaque but very narrow discs, optically thin dust (where unshielded PR drag occurs), a pure gas disc or discs that have been fully consumed (accretion has ceased). In this work we explore the latter argument, in terms of a shorter disc lifetime resulting from more rapidly-driven accretion. However, we note that this explanation cannot apply to polluted warm DAs with no observable discs, as the short sinking timescales of these WDs mean they must be actively accreting material at the time of observation. Other studies have also proposed mechanisms to shorten the disc lifetime. \citet{jura_2008} suggested that collisions between multiple small asteroids can create gas, which could increase the accretion rate of the dust due to viscosity. It has also been suggested that a build up of gas in the disc could cause runaway accretion which enhances the accretion rate and decreases the disc lifetime \citep{rafikov_2011b, metzger_rafikov_2012}.

As most polluted WDs have no or unobservable discs, the accretion process is usually explored indirectly. In systems containing warm DAs (with effective temperatures $\ge$ 10\,000K), where short sinking timescales imply active accretion, it is assumed that the system is in steady state and the rate of accretion is estimated from the the elemental sinking timescales and their observed abundances \citep{dupuis_fontaine_1992, gansicke2012chemical, farihi_2016}. Because the sinking timescales are short compared to the accretion timescale, this offers a robust lower estimate on the accretion rate: i.e.\ the minimum amount of mass needed to replenish the pollutants in the atmosphere before they become undetectable.
In the case of the cooler DAs and DBs, it is more difficult to determine if the WD is actively accreting. The sinking timescales in these systems are much longer, meaning pollutants can be present after the disc has been depleted. Assuming the pollution abundances follow the disc composition (typically assumed to be bulk terrestrial mass) while it is accreting, it is possible to use the ratio of elemental abundances to determine if the system is actively accreting: i.e.\ the system is accreting if the abundance ratios match the expected disc composition \citep{jura_farihi_2007,  zuckerman2010, farihi_gan2012}. 
For scenarios where accretion has stopped, the steady state approximation is used to infer the past accretion rate. That is, the observed abundances can be used to infer the mass of the pollutants in the convection layer and are integrated with their sinking timescales to derive a `historical' accretion rate. In most cases, the accretion timescales are much longer than the diffusion timescales, meaning the steady state approximation can be considered safe \citep{koester_2009}. However, for the systems we consider in this paper (typically DAZ and DZ $<$ 10\,000K) the accretion timescales are comparable to, or smaller than, the sinking timescales. This introduces uncertainty to the predicted accretion rates, notably for cooler WDs, but estimates are expected to be accurate within an order of magnitude \citep{farihi_jura_2009, farihi_2016}.

A study by \cite{bauer_bildsten_2018} builds on the work of \cite{deal_deheuvels_2013} and \cite{wachlin_vauclair_2017}, and highlights that the inclusion of thermohaline mixing requires higher accretion rates in polluted WDs. Thermohaline mixing results in turbulence in convective layers of WDs, specifically DA type, caused by the higher mean molecular weight of metals sitting above the hydrogen layer. The sinking of metals into the WD atmosphere is normally thought to be controlled by gravitational settling alone, however the inclusion of thermohaline `fingering' increases the mixing of metals below the surface, which decreases the observable surface metals more rapidly. This would require the accretion rates to be orders of magnitude higher than those inferred from gravitational settling alone. A rebuttal in \cite{koester_2015} to the initial \cite{deal_deheuvels_2013} study suggested that thermohaline mixing is likely to be negligible due to the uncertainties in the efficiency of mixing and the small abundances of the heavy elements. However, studies using thermohaline mixing find a better match between the observed metal abundance ratios to the accreted material, assuming bulk Earth composition. If these studies are correct then the accretion rates may be orders of magnitude above the predictions of PR drag and gravitational settling alone. Other studies modelling convective overshoot in WD atmospheres find that accretion rates can be underestimated by up to an order of magnitude \citep{cunningham2019}.

When an object is tidally disrupted, the resultant debris stream will initially follow the original orbit of the asteroid \citep{jura_2003}. \citet{nixon_pringle_2020} propose that these eccentric orbits can explain the variability in dusty debris and \citet{vanderbosch_hermes_2019} observe an eccentric stream of debris orbiting ZTF J013906.17+524536.89. However, the majority of observations of discs appear to show the debris on a near circular orbit inside the tidal radius \cite[See][and references therein for full review]{farihi_2016}.
The transition from an eccentric, tidally disrupted debris stream to circular dust disc is not well understood. Currently, PR drag is employed as the main method of circularising the disc, but the timescales for this are thought to be approximately Myrs \citep{farihi_jura_2009, rafikov_2011a}.
There have been some suggestions of how these discs might circularise faster, e.g. precession causing collisions \citep{veras_leinhardt_2014} and dust-gas coupling \citep{bonsor_farihi_2017}. In this paper, we explore the idea that a magnetic field anchored on the white dwarf could induce a drag force on the debris particles, circularising their orbits more rapidly. 

Following the Sloan Digital Sky Survey (SDSS), the number of known WDs is now in the 10,000s. Around 2-20\% of these WDs have been observed to show some evidence of magnetism \citep{ferrario_demarco_2015}. Assuming a standard distribution and no causal relationship between magnetism and pollution, there should be a subset of WDs that are magnetic and polluted. Indeed, in the SDSS data release 12 (DR12) 6,576 new WDs and sub dwarfs were identified, 315 of which are polluted WDs and 37 are magnetic WDs. Of these, 7 are classified as DZH, i.e.\ both polluted and magnetic \citep{sdss_dr12}. 
There appears to be a higher incidence of magnetism in cool polluted white dwarfs which might indicate a correlation between the two \citep{kawka_vennes_2014, hollands_gainsicke_2015, kawka_vennes_2019}.
There is also a population of cataclysmic variables where approximately 25\% of the WDs are magnetic \citep{ferrario_demarco_2015}. It has long been known that a magnetic field on a WD (or other accreting body) can have a dramatic effect on accretion dynamics \citep[e.g:][]{cropper_1990, collier_campbell_1993, king_1993, wynn_king_1995, schmidt_hoard_1999, li_wilson_1999, Wu_2000, alecian_wade_2007, bouvier_alencar_2007, gregory_donati_2011, zhilkin_bisikalo_2012, hussain_alecian_2014, wickramasinghe_2014,  isakova_zhilkin_2017, vonboxsom_2018, ablimit_maeda_2019, shahbaz_2019}.  
\citet{farihi_vonHippel_2017} use a model where dust is trapped in the magnetosphere of a WD to explain the observational anomaly in WD1145+017. The authors conclude that the interaction between dust and a magnetic field is likely to be important in explaining the dynamics of this particular system and estimate the distance the magnetic field begins to impact the dynamics. Other recent studies have looked at the effect of magnetic white dwarfs on iron core asteroids, demonstrating that Ohmic heating and Lorentz drift have a significant influence on dense, large, ferromagnetic objects \citep{veras_wolszan_2019, bromley_kenyon_2019}. In contrast, following earlier studies by \cite{king_1993, wynn_king_1995, wynn_king_1997, meintjes_2005, wickramasinghe_2014}, we examine the effect of diamagnetic drag forces on small debris fragments.

In this paper we examine how a magnetic field may effect the evolution of a dust disc around a polluted white dwarf. Section \ref{sec:num_calc} establishes a parameterised model of the field-dust interaction. Sections \ref{sec:params} and \ref{sec:Part_Orb_Calc} then detail the parameter range for WDs and tests orbital calculations against those from \citet{veras_leinhardt_2014}. Section \ref{sec:results} presents the results of the magnetic model which we analyse and discuss in section \ref{sec:DISS}. Finally, we conclude this work in Section \ref{sec:Conclusion}.

\medskip
\section{A Model for the interaction of a debris disc and a magnetic field}\label{sec:num_calc}
\subsection{Physical Motivation}

Studies of the abundances of volatiles in the atmospheres of polluted WDs indicates that asteroids are made up of lots of different materials, some of which are paramagnetic (Ca, Mg, Na, O, Fe, Ni, Al) and some diamagnetic (C, N, Si, H$_2$0). The abundances observed show that the majority of the asteroid is formed of diamagnetic elements \citep{koester_gansicke_2014, harrison_bonsor_2018, swan_farihi_2019b}.
All materials are either ferromagnetic, paramagnetic or diamagnetic. 
Diamagnetism is a quantum mechanical effect present in all materials, but is usually weak and only manifests itself in the absence of paramagnetism and ferromagnetism. Superconductors are good examples of strong diamagnets, as they exclude magnetic fields from their interior entirely, known as the Meissner effect \citep{bardeen1957}. 

A diamagnetic object (or conducting object) passing through an applied magnetic field will experience a force caused by the generation of an oppositely-aligned internal field, preventing the external field passing through and resulting in a force on the object as the field lines warp around it. This force acts to oppose the motion of the object across the magnetic field lines  \citep{berry1997, kustler2007}. The magnitude and direction of the resulting drag force is dependant on the relative velocity between the object and the external field \citep[e.g:][]{drell_foley_1965}, and the structure and composition of the object's constituent material. Previous studies  \citep[e.g.][]{king_1993, king_regev_1994, wynn_king_1995, ultchin_regev_1997,  ultchin_2002,  norton_wynn_2004, norton_butters_2008, wickramasinghe_2014} have considered the effect of a magnetic field on weakly-conducting, diamagnetic gas in magnetic cataclysmic variable and T-Tauri stars. Other work has pointed out that similar forces may also be imparted on diamagnetic dust particles via interaction with plasma, or stellar irradiation \citep{fuerbacher_fitton_1972, grun_morfill_1984}.

\subsection{Dust Parameters}
Models of tidal disruption radii find that particle densities between 1-3.5 g cm$^{-3}$ yield disruption radii between 0.8 -- 1.2 R$_{\odot}$, which best fit observation \citep{farihi_2016}. Observations of hundreds of Solar system asteroids give density estimations between 1--8 g cm$^{-3}$, with the majority in the 1--3 g cm$^{-3}$ range \citep{michalak_2000, carry_2012, hanus_viikinkoski_2017}.

In terms of particle size, nanometre size dust should be quickly removed from the system by radiative forces as PR drag causes these particles to spiral in and sublimate before being blown out of the system \citep{burns_lamy_1979}. Hence, we expect most particles interacting with the WD magnetic field to be micron sized and, indeed, these form the bulk of the population of the observed dust in polluted WDs with discs \citep{jura_2003, jura_farihi_2007, farihi_2016}. 

\subsection{The Magnetic Drag Force}\label{sec:analyticeqns}

Given that the highest abundance of material in rocky bodies is diamagnetic, we parameterise the force felt on a diamagnetic dust particle due to the magnetic field following \cite{king_1993, wynn_king_1995} as 
\begin{equation}
F=-m_p kv_{r},
\end{equation}
where $m_p$ is the particle mass, $v_r$ is the relative velocity between the local magnetic field and the particle and $k$ is a drag coefficient, which is determined by the local magnetic field strength and particle size, composition, and charge
(see e.g.\ \cite{ghosh_lamb_1979, grun_morfill_1984, gruen_gustafson_1994,chancia_hedman_2019, lhotka_2019}).
When the local magnetic field lines are moving more slowly than the particle ($v_r > 0$), the particle feels a force that opposes its motion. The force acts on the particle motion perpendicular to the field lines,  causing it to lose energy and angular momentum and, eventually, to accrete onto the WD surface. If the field lines are moving faster than the particle ($v_r < 0$), the particle gains energy and angular momentum, causing a net outward motion from the star. 

The magnetic tension force, caused by the field lines warping around the diamagnetic dust particle, is given by 
\begin{equation}
F \simeq \frac{V}{R_c}\frac{B(r)^{2}}{8 \pi},
\end{equation}
where $V$ is the particle volume, $B(r)$ is the local magnetic field strength and $R_c$ is radius of field line curvature. We assume $R_c$ to be approximately the size of the particle ($R_p$) and that 
$V\sim R_p^{3}$, giving
\begin{equation}
F \sim R_{p}^{2}\frac{B(r)^{2}}{8 \pi}.
\end{equation}
The magnitude of torque on the particle ($\boldsymbol{\tau}$) exerted by the magnetic field  is
\begin{equation}
|\boldsymbol{\tau}| = |\mathbfit{r} \times \mathbfit{F}| \sim R_{p}^{2} \: r\frac{B(r)^{2}}{8 \pi},
\label{eq:tau}
\end{equation} 
where $\mathbfit{r}$ is the position vector of the particle from the centre of the WD. 
The angular momentum loss timescale of a particle can be estimated as 
\begin{equation}
    T_{L} \sim \frac{|\mathbfit{J}|}{|\boldsymbol{\tau}|} 
    \label{eq: j/tau} 
\end{equation}
where $\mathbfit{J}$ is particle's orbital angular momentum and
\begin{equation}
|\mathbfit{J}|={m_{p}(GM_{WD}r)^{1/2}},
\label{eq:J}
\end{equation} 
assuming a circular Keplerian orbit. 
\noindent Here $m_p$ and $M_{WD}$ are the masses of the particle and white dwarf respectively. Using equations \ref{eq:J} and \ref{eq:tau}, we can estimate a typical particle lifetime, or accretion timescale
\begin{equation}
T_{L} \sim \frac{m_p}{R_{p}^{2}}\frac{8 \pi}{B(r)^{2}}\left(\frac{GM_{WD}}{r}\right)^{1/2}
\sim \rho R_{p} \frac{8 \pi}{B(r)^{2}}\left(\frac{GM_{WD}}{r}\right)^{1/2},
\label{eq: Life}
\end{equation}
where $\rho \sim m_p R_p^{-3}$ is the particle density.  

The drag parameter $k(r)$ is related to the particle lifetime via 
\begin{equation}
k(r) \sim \frac{1}{T_{L}}.
\label{eq: k_r_T_L}
\end{equation}
At the surface of the WD 
\begin{equation}
k(R_{\star}) \simeq k_0 \sim \left[\rho R_p \frac{8 \pi}{B_{\star}^{2}}\left(\frac{GM_{WD}}{R_{\star}}\right)^{1/2}\right]^{-1}, 
\label{eq: k_0}
\end{equation}
where the subscript $\star$ indicates values at the WD surface. 
We assume the large scale structure of the magnetic field to be dipolar and aligned with the spin axis of the WD, such that $k(r)$ scales as
\begin{equation}
k(r)= k_0 \left(\frac{r}{R_{\star}}\right)^{-6}\left(1+3\frac{z_p^2}{r^2}\right)^{1/2}
\end{equation}
where $z_p$ represents the particle position along the spin/dipole axis. In this work we further assume the orbital plane of the particles to be perpendicular to the spin axis, such that $z_p$ will always be 0, yielding the simple scaling
\begin{equation}
\label{eq:Kr}
k(r)= k_0 \left(\frac{r}{R_{\star}}\right)^{-6}.
\end{equation}

\subsubsection{Magnetic Drag Force Compared to PR drag}
To determine the radius at which magnetic drag dominates PR drag, we equate the maximal effects of unshielded PR drag \citep{robertson_1937, burns_lamy_1979} and the magnetic drag force (assuming a circular orbit, where the Mie Scattering coefficient, $Q_{PR}$, equals unity)
\begin{equation}
R_p^2 \frac{B_\star^2}{8\pi} \left(\frac{R_\star}{r}\right)^6 = \frac{R_P^2 R_\star^2\sigma T^4}{r^2c}.
\label{eq:force1force}
\end{equation}
Solving for distance ($r$) produces an estimate of the radius within which magnetic effects are expected to be dominant (relative to unshielded PR drag), analogous to Alfven radius adopted in other magnetic accreting systems.
\begin{equation}
R_{N} = \left(\frac{B_\star^2 c R_\star^4}{8\pi \sigma T^4}\right)^{1/4}.
\label{eq:Nevfla}
\end{equation}
$R_N$ represents the radius of magnetic dominance for a micron sized particle over unshielded PR force for a 10\,000K WD. Shielded particles, in contrast, only experience a fraction of the incident radiation and the equivalent magnetic dominance radius will be larger. 

\begin{figure}
\centering
\includegraphics[width=\columnwidth]{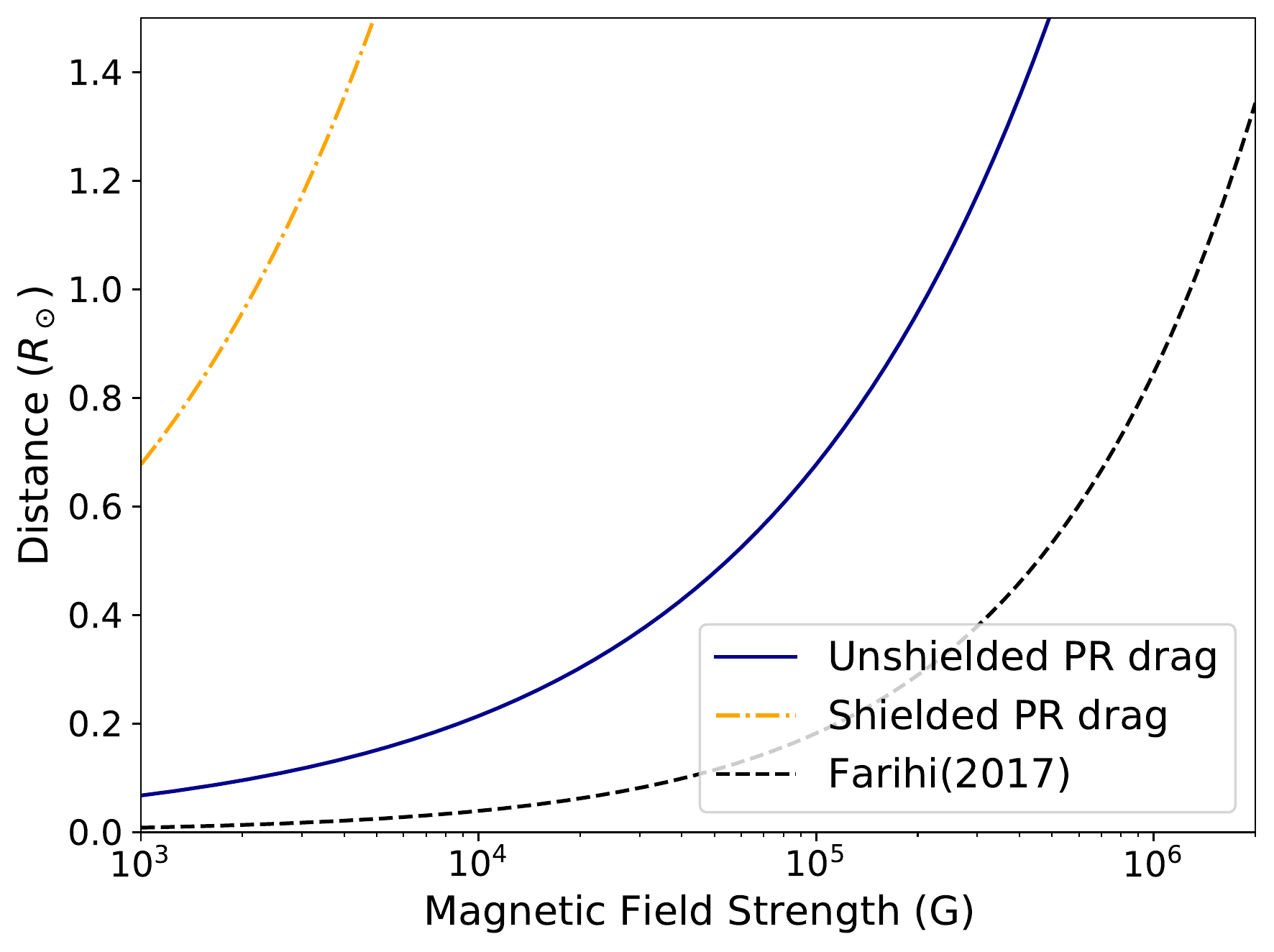}
\caption{\label{fig:Rmag_B} Distance of magnetic dominance for a 10\,000K WD from equation \ref{eq:Nevfla} compared to the distance determined by equation [9] from \citet{farihi_vonHippel_2017}}.
\end{figure}

We compare equation \ref{eq:Nevfla} with equation [9] from \cite{farihi_vonHippel_2017}, which estimates the distance at which the magnetic field will influence the dust, in Figure \ref{fig:Rmag_B}.
For field strengths in the 100kG regime or greater, the magnetic field will dominate throughout the region where circularised debris discs are typically found. 

To estimate the region of magnetic dominance for a shielded disc, we estimate the ratio of the expected lifetimes of an unshielded disc ($\sim$ 100 years) and a shielded disc ($\sim$ million years) as $\sim10^{-4}$. This can be applied as a parameter to equation \ref{eq:force1force}, giving a rough estimate of 5kG where magnetic drag is dominant within the tidal disruption radius. This gives us two limits on the fields at which we estimate  magnetic drag may dominate PR drag within the tidal disruption radius: 100kG for unshielded PR drag, and 5kG for shielded PR drag. 

For fields of order 1-10 kG, the drag on the unshielded debris will be influenced by the magnetic field as it approaches the WD due to the $r^{-6}$ dependence of the magnetic drag force.
Eccentric discs are expected to have closest approaches within the tidal disruption radius, highlighting the magnetic drag force as a mechanism for circularisation for an eccentric disc. Eccentric discs forming in weaker fields, less than 10kG, will follow the PR drag path except at closest approach where the field may accelerate the circularisation of the disc if the pericentre is close enough to feel the magnetic effects. If the magnetic field is strong enough, the drag on the dust could cause it to break, creating a polar funnel \citep{krzeminski_serkowski_1977, wickramasinghe_2014}. Whether this happens depends on where the dust is in its orbit. This introduces the critical radii that need to be considered when looking at the effects of a magnetic field.

The tidal disruption radius is the point where a body would be destroyed by the tidal stresses imparted by the star. This radius differs depending on the density and tensile strength of the body, but mostly sits at around one solar radius for the average rocky body around a white dwarf ($R_{TD} = 1R_\odot$).

The co-rotation radius is the distance at which the velocity of the particles orbiting the WD is the same as the spin of the stellar field.
The co-rotation radius for a circular orbit is given by:
\begin{equation}
R_{co} = \left(\frac{GM_{WD}}{\Omega_{mag}^2}\right)^{1/3}.
\label{eq:CO-RO}
\end{equation}
Where $\Omega_{mag}$ is the angular velocity of the magnetic field, which is equivalent to the WD's angular velocity because the field is fixed to the stellar surface. 
\\
 
Using the three radii; the tidal disruption radius $R_{TD}$, co-rotation radius $R_{co}$ and magnetic dominance radius $R_N$, we show their importance and how they relate to physical outcomes: 
\begin{itemize}
  \item If $R_N \ll R_{TD}$, (i.e. if the magnetic drag force is much smaller than unshielded PR-drag) magnetic effects are expected to be negligible on the evolution of the disc. This is estimated as the point where PR drag dominates down to 0.15$R_\odot$. For a WD of 10\,000K, the field becomes negligible at $\lesssim$ 5kG for an unshielded disc and $\lesssim$ 500G for a shielded disc. \\  
  \item If $R_N \ge R_{TD}$, the diamagnetic drag term is expected to dominate in the disc forming region. We predict this to be the case for field strengths $\ge$100kG for an unshielded disc and $\ge$5kG for a shielded disc. \\
  \item For $R_N \ge R_{co}$, the particles will get a net gain in angular momentum as the field lines are moving faster than the particles orbit. Either trapping the particles, pushing them into a wider orbit, or in extreme cases ejecting them entirely. This is assuming the particle orbit is the same direction as the stellar spin. Typical single white dwarf spins vary from a few hours to days 
  \citep{hermes2017}. Assuming a spin of 2 days, $R_{co} = 5.64R_\odot$. This means magnetic trapping and diamagnetic propeller systems should be considered for field strengths $\gtrsim$ 5MG for an unshielded disc and $\gtrsim$100kG for a shielded disc.  
  \\
  \item For $R_N < R_{co}$, the particles lose angular momentum passing through the field lines.\\
\end{itemize}

\noindent For the purposes of this paper, we assume that that $R_N < R_{co}$, equivalent to a slow spinning, average mass, moderately magnetic white dwarf.

\subsection{Particle Orbit Calculations}\label{sec:5_addingforce} 

In this section we describe the process of adding a diamagnetic drag force into \texttt{Freefall}\footnote{\href{https://github.com/Ry-C123/Freefall}{Github Link}  \citep{cutter_Hogg2020}}, an orbital integrator we use to study the evolution of an eccentric disc over time.


Using the formalism from \citet{ultchin_2002}, we show that the force term can be expressed proportionally to the velocity component perpendicular to magnetic field $\mathbfit{F}=-k\mathbfit{v}_{\perp}$. Where $k$ is equal to equation \ref{eq:Kr} multiplied by $m_p$ and $\mathbfit{v}_{\perp}$ is the velocity component perpendicular to the magnetic field.

A particle moving into a magnetic field gains an induced current, described by Faraday's induction law. According to Lenz's laws the induced current creates a magnetic field around the particle that opposes the field of the magnetic field proportional to its velocity. This can also be called magnetic breaking. Magnetic breaking opposes the velocity of the Lorentz force, the unit vector of this force can be expressed:


\begin{equation}
    \mathbf{\hat{F}} \propto \int \mathbf{\hat{J}} \times \mathbf{\hat{b}}  d\tau .
\end{equation}
Where $\mathbf{\hat{J}}$ is the local current density, $\mathbf{\hat{b}}$ is the local field line direction combined with Ohms law and the Lorentz force we get:

\begin{equation}
\mathbfit{F} = -k\mathbf{\hat{b}} \times (\mathbf{\hat{b}} \times \mathbfit{v}_r) .
\label{eq:Bunit}
\end{equation}
As $\mathbf{\hat{b}}$ is a unit vector, \ref{eq:Bunit} can also be represented as:

\begin{equation}
\mathbfit{F} = -k\left(\mathbfit{v}_r - \mathbf{\hat{b}}(\mathbf{\hat{b}} \cdot \mathbfit{v}_r)\right) .
\end{equation}

Where $v_r$ is the relative velocity. This relative velocity can be surmised as the difference between the rotational velocity of the magnetic field and the orbital velocity of the particle. This relative term is used to find the force component in the frame of the rotating field:
\begin{equation}
\mathbfit{v}_r = \mathbfit{v}_{particle} - \mathbfit{v}_{field} .
\label{eq:relV}
\end{equation}

We convert to acceleration and find the change in velocity for a particle in a magnetic field by multiplying by the time step.
\begin{equation}
\delta \mathbfit{v}_p = -k\frac{\left(\mathbfit{v}_r - \mathbf{\hat{b}}(\mathbf{\hat{b}} \cdot \mathbfit{v}_r)\right)}{m_p} dt .
\end{equation}
This force is now integrated into the \texttt{Freefall} architecture.

The force felt by the particles changes as $r^{-6}$ which creates a steep gradient in the force at high magnetic field strengths and close distances. This makes the simulation 'stiff' at small distances from the WD. To remedy this, we include a softening length, $\epsilon$, to flatten the gradient at close approach. Field strengths $<$1MG do not need a softening parameter, $\ge$1MG fields has an epsilon of 0.15$R_{\odot}$, $\ge$10MG has a epsilon of 0.2 $R_{\odot}$, and anything $\ge$ 100MG an epsilon of 0.8$R_{\odot}$. This alters equation \ref{eq:Kr} to:

\begin{equation}
k(r)= k_0 \left(\frac{\sqrt{r^2+\epsilon^2}}{R_{\star}}\right)^{-6} .
\label{eq: Kr2}
\end{equation}

\section{Parameter Range of WD}\label{sec:params}

\subsection{Mass and Radius}
The WD mass distribution is double peaked and has a lower limit at 0.2$M_{\odot}$ and an upper limit at 1.4$M_{\odot}$ which is known as the Chandrasekhar limit \citep{chandrasekher_1931}. Most WDs are around 0.6$M_{\odot}$; however, magnetic WDs tend to typically be slightly higher in mass at approximately 0.8$M_{\odot}$ \citep{ferrario_demarco_2015}. This is likely due to their formation mechanism which is believed to involve binary evolution or formation from previously high mass magnetic A and B type stars \citep{wickramasinghe_ferrario_2000, tout_wickramasinghe_2008, nordhaus_2011, briggs_ferrario_2015, briggs_ferrario_2018}. For simplicity we will use a typical WD mass of 0.6$M_{\odot}$.

We estimate the radius of the WD using the mass-radius relation of \cite{nauenberg_1972}. For a 0.6$M_{\odot}$ WD the radius is calculated to be 0.0126$R_{\odot}$, for the typical magnetic WD at 0.8$M_{\odot}$ the radius is 0.0101$R_{\odot}$.
The surface gravity is high in these objects because of the mass-radius relation and is found to be between 6.50-9 in log $g$ space. A typical WD is expected to have a surface gravity of log $g$=8.0. It is standard practice to assume a surface log $g$ of 8.0 if the gravity is not known \citep{kepler_romero_2017, hollands_gansicke_2018}.

\subsection{Spin rate}
The spin of magnetic WDs tends to be on the order of years, most likely due to magnetic breaking \citep{fontaine_and_brassard_2008}. The upper limit for spin for single WDs is around 700--300 seconds while the longest measured spin rate is hundreds of years \citep{barstow1995, jordan2002, reding2020isolated}. There is some evidence of a bimodal distribution of spin rates with a small peak at the spin rate of hours, which is thought to be caused by mergers and accretion events which spin up the star, and a larger peak at years  \citep{wickramasinghe_ferrario_2000, garcia-berro_kilic_2016}. Non-magnetic WDs similarly have fairly slow rotation periods $>1hr$ and the mean period of a non-magnetic WD is 1 day \citep{wickramasinghe_ferrario_2000, kepler_romero_2017}. 
Here we study the effects of a magnetic field alone and the spin of the star can effect the dynamics around the WD drastically; so, we use a non-rotating WD. The effects of different spins will be studied in future work. \\ 

\subsection{Magnetic Field Strength}
The most magnetic WDs are observed to have field strengths of the order MG, with the highest observed magnetic field strength at 800MG \citep{ferrario_demarco_2015}. 

\begin{figure}
    \centering
    \includegraphics[width=\columnwidth]{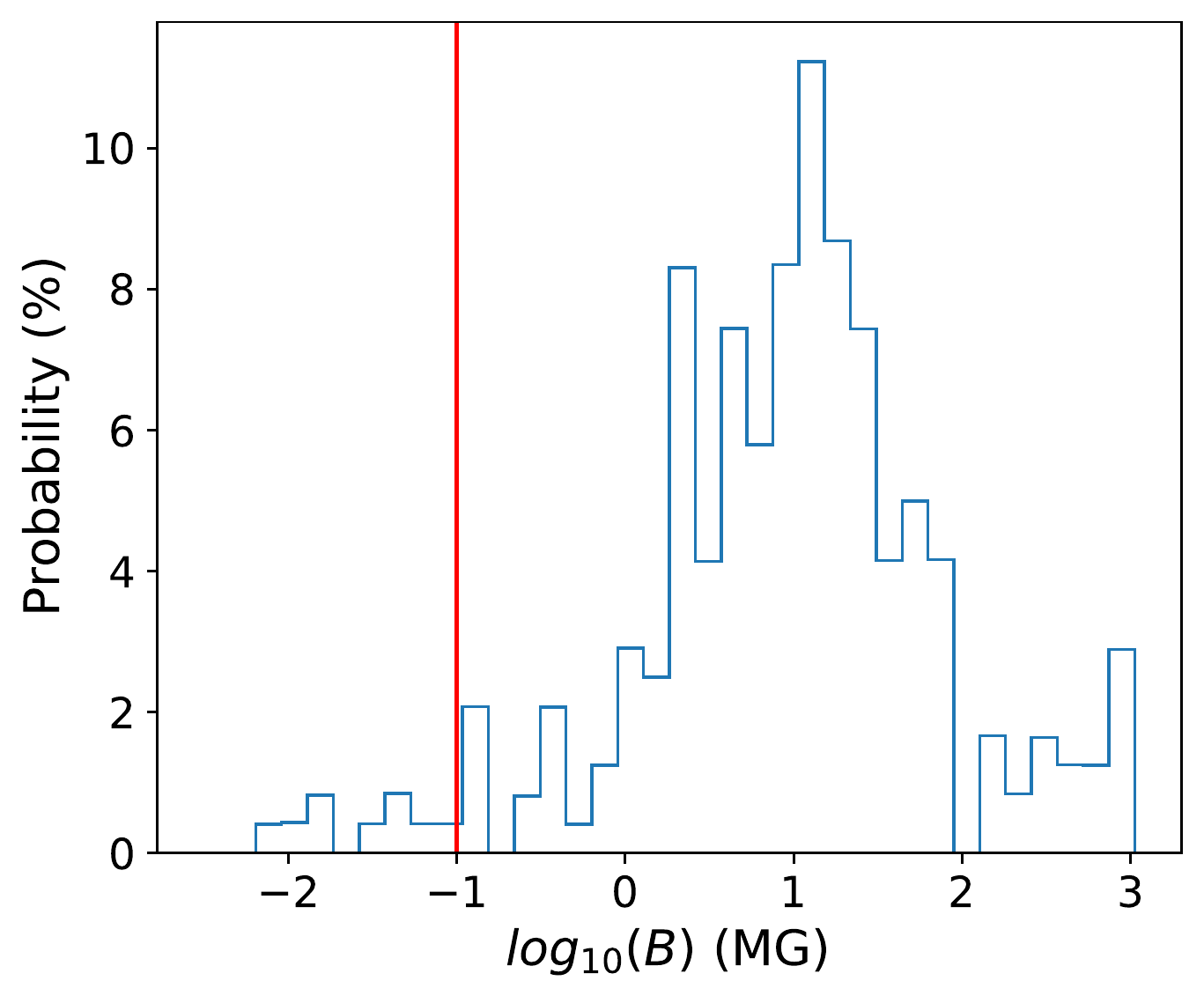}
    \caption{\label{fig:distribution}A probability distribution of magnetic WDs by field strength based on \citet{ferrario_demarco_2015}. The vertical line indicates 100kG in field strength, the expected point where magnetic effect becomes dominant compared to unshielded PR drag.}
\end{figure}

Magnetic fields are observable down to tens of kilo Gauss, below this the Zeeman splitting goes beneath the resolution of the spectra, meaning field strengths cannot be reliably determined. Due to this, the true lower limit for magnetic fields in WDs is unknown. 
A study of 170 magnetic WDs finds a the most common magnetic field strength peaks at 1MG (see Figure \ref{fig:distribution}, It is estimated that between 2-20\% of WDs are magnetic \citet{ferrario_demarco_2015}). 

The handful of polluted magnetic WDs that have been observed have field strengths of 50kG up to 10MG \citep{kawka_vennes_2014}, which is the parameter space where magnetic influence is expected.

We investigate a variety of different magnetic field strengths which cover the range of observed strengths and some below the observable limit: 1KG, 10KG, 100KG, 1MG, 10MG, and 100MG.

\subsection{Temperature}
WD temperatures can vary from 150\,000K, when they originally form, to $<$6\,000K, WDs that have been cooling for billions of years.
We do not expect the magnetic field to be effected by the temperature of the WD. The majority of polluted WDs we have observed have temperatures $<$25\,000K, but the incidence of magnetism is quite common in low temperature WDs $<$8\,000K 
\citep{kawka_vennes_2014, hollands_gainsicke_2015, kawka_2018}; although this could be a result of small number statistics. 

As the influence of the magnetic field drag is considered independent of thermal properties, PR drag becomes the only drag term affected by the white dwarf temperature. We set 10\,000K (which equates to a cooling age of 0.5Gyr) as a typical WD temperature.

\section{Particle Orbit Calculations} \label{sec:Part_Orb_Calc}
\subsection{Parameters of Polluting Particles} \label{param_asteroids}
To simulate the typical case of a pollution event we also require some information about the body that will be tidally destroyed. We build up a rubble-pile asteroid made of hard spherical micron sized particles as they are the most abundant in the observations \citep{jura_2003, jura_farihi_2007, farihi_2016}. Our asteroids will therefore be relatively small in terms of mass as it would require a $10^{14}$ particles to reach a $100$kg sized asteroid. We use a relatively small number of particles (500) to find the time it takes for the disrupted asteroid to circularise and accrete onto the WD surface. Figure \ref{fig:diff_part_num} in the appendix highlights that the disc lifetime is not highly effected by particle numbers so we can use a small number of particles to run the simulations faster. As we are using a small number of micron sized particles we expect that collisions would not play a large role, even after circularisation, but studies with larger numbers of particles would likely require them \citep{farihi_2008b}. Similarly the lack of particles means that we would not expect particles to shield each other, and therefore expect the particles to feel the full effect of the incident starlight, as in the unshielded PR drag regime of \citet{burns_lamy_1979}.
We assume a dust particle density of 2$gcm^{-3}$ and radius of $10^{-6}$m. 

\subsection{Orbital Integrator Initial Test}\label{sec:4_nbodycode}
The use of a collision-less simulation means that a full N-Body code is not required as the debris particles are not interacting. We are therefore able to use a simplified orbital dynamics module where the only forces on each particle are gravity (including a GR correction), unshielded PR drag, and the drag from the magnetic field. We employ \texttt{Freefall}, which was created specifically for this work. For our simulations, we use a 4th order Runge-Kutta (RK) method. It should be noted, as the magnetic force is dependant on velocity, leapfrog methods will become unstable at high accelerations. This is primarily why we are using RK methods.     

As a comparison we first recreate the results of  \citet{veras_leinhardt_2014}, and use the results of that simulation as our initial conditions and then add in the PR and magnetic forces to see how the disc evolves.
To do this we use the same values to set up the simulation and compare our results in a non magnetic scenario. We use the same initial parameters for mass and radius of the WD ($M_{WD}$, $r_{WD}$) and small body ($M_A$, $R_A$) as well as the initial eccentricity ($e$) and major semi axis ($a$). 

We use the following parameters taken from \citet{veras_leinhardt_2014}:
\begin{itemize}
    \item $R_{A}$= $3 km$
    \item $a=0.2AU$
    \item $e=0.9966$
\end{itemize}

\citet{veras_leinhardt_2014} uses the semi-major axis of 0.2AU as it is computationally less intensive. 


\texttt{Freefall} does not have a provision for sticking bodies together; instead, we model the orbit of the asteroid and take the orbital data as the asteroid passes the disruption radius ($\approx$ 1$R_\odot$) and insert test particles with the same parameters and a Gaussian distribution of the particles in the X, Y, and Z directions to model a 3km asteroid. To mimic a disruption of the asteroid we also insert a Gaussian perturbation in the velocity of the particles in the orbital plane. 

We use the same number (5000) of $4.52\times10^{11}$kg mass particles so the simulations match as closely as possible. We are able to keep the same computational stability as PKDGRAV using an 8 second integration time step. By using similar input parameters for the WD and asteroid, we are able to recreate the results of \citet{veras_leinhardt_2014} with a high degree of accuracy. We were able to run this simulation in days as apposed to the month timescale for the simulations using PKDGRAV \citep{stadel2002high} with the same accuracy (shown in the appendix \ref{fig:filling}).
\medskip

\section{Results}\label{sec:results}
We first take the analytic equations derived in section \ref{sec:analyticeqns} to study the lifetimes of circularised discs for different particle sizes, field strengths and orbital distances. We then use the equations from section \ref{sec:5_addingforce} for our simulations to see how closely they align with the analytic equations. We finally use the simulations to see how an eccentric disc circularises and its lifetime due to magnetic fields of different strengths compared to unshielded PR drag alone. 
\subsection{Analytic Results} 

\begin{figure}
\centering
\includegraphics[width=\columnwidth]{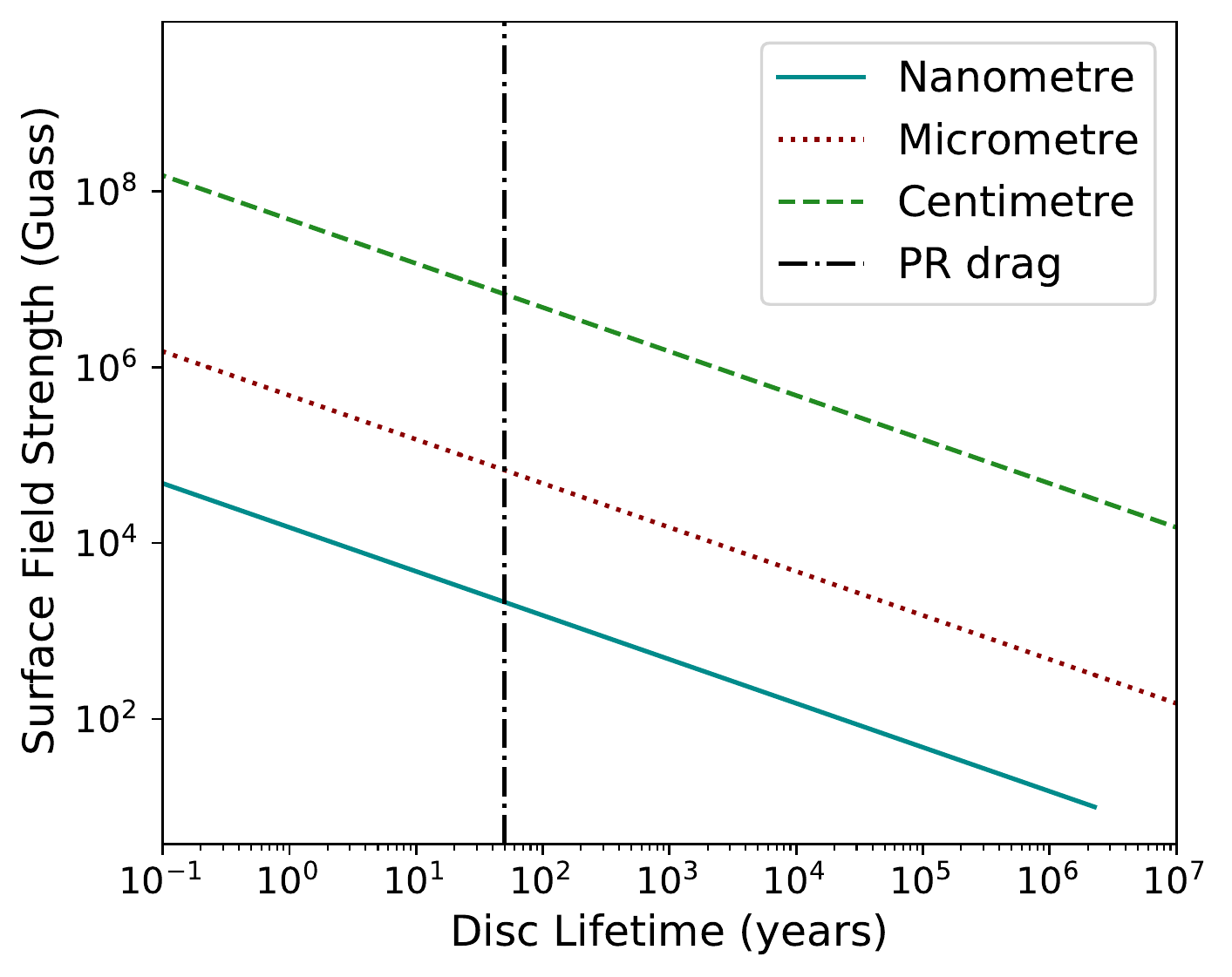}
\caption{\label{fig:magfield_disclife} Disc lifetime for nano, micro, and centimeter size dust grains starting at 1 solar radii for varying surface magnetic field strengths. The black dashed line indicates the lifetime of the particle due to unshielded PR drag of a 10,000K WD.}
\end{figure}

Using equation \ref{eq: Life} for nano, micro and centimetre sized dust with  $\rho \sim 2$ g cm$^{-3}$, we can estimate the expected lifetimes of these particles for different WD field strengths. Figure \ref{fig:magfield_disclife} shows the lifetime estimates for typical values $M_{WD} = 0.6  M_\odot $, $R_{\star} = 0.0126  R_\odot $ and $ r= 1  R_\odot $, with this latter being the particle's orbital radius (equivalent to the disc radius).
From Fig \ref{fig:magfield_disclife} we see that moderate WD surface fields of kilo-Gauss are able to reduce the disc lifetime (accretion timescale) below that expected from shielded PR drag (where the lifetimes of an unshielded micron sized particle is $\sim$ 50 years and shielded particle is $>10^5$ years) \citep{rafikov_2011a, farihi_jura_2009, farihi_2016}. 

\begin{figure} \centering
\includegraphics[width=\columnwidth]{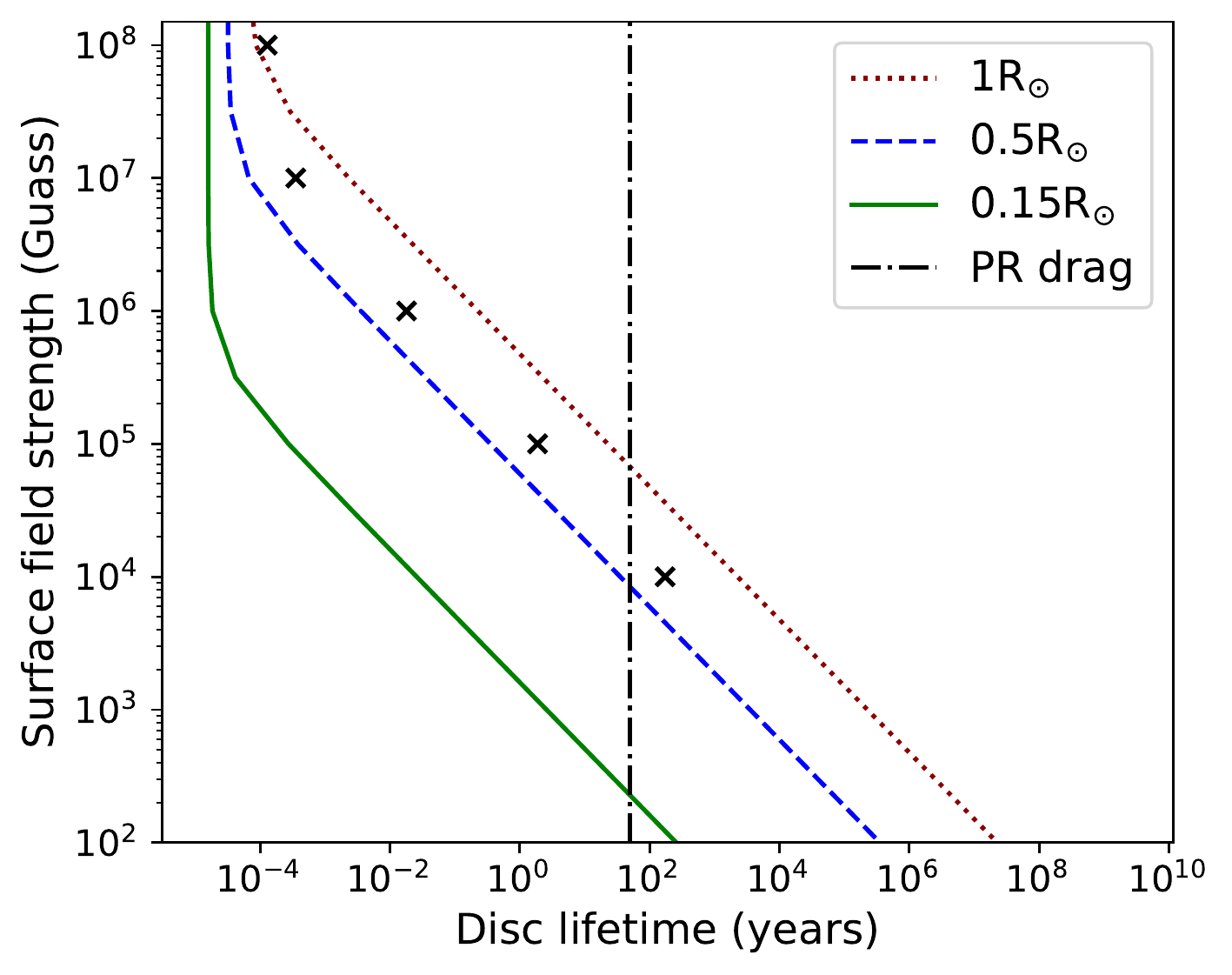}
\caption{\label{fig:Tlife_tets} Simulated lifetime of a micrometer dust grain starting at 1, 0.5, and 0.15 solar radii in a circular orbit. The lines represent the predicted lifetime of a micron particle from a magnetic field (figure \ref{fig:magfield_disclife}). The vertical dashed line indicates the lifetime of the particle due to unshielded PR drag of a 10,000K WD. The black crosses show the results of simulations at various B field strengths without the inclusion of PR drag.}
\end{figure}

Figure \ref{fig:Tlife_tets} shows lifetime estimates at different orbital distances for a micron sized particle. At the shortest timescales, the particles fall in to the WD on a dynamical timescale. 

\subsection{Results of Single Particle Test}\label{sec:6_spt_summery} 
To compare our simulations with our analytical results we set up a simulation of a single particle at 1 solar radius and find the time taken for the particle to get to 2 WD radii (0.02$R_{\odot}$), at which point we consider it accreted. We do this for various field strengths as well as the non magnetic case to test the effect of unshielded PR drag alone. We find that the lifetime of the particle due to these magnetic changes are much shorter than the timescales presented by PR drag alone, highlighted in Figure \ref{fig:Tlife_tets}.

We expect the simulated disc lifetime to be shorter than what is predicted by equation \ref{eq: Life} for a few reasons. Firstly, the analytic equation assumes a constant torque throughout the particles lifetime. In reality, as the particle approaches the white dwarf, the torque increases which shortens the path length. Secondly, the accretion radius is at 2 WD radii rather than 1. This means the lifetime is cut short for particles that have not yet lost all of their angular momentum. 
We anticipated from equation \ref{eq:Nevfla} that magnetic drag would dominate PR drag at 100kG. Figure \ref{fig:Tlife_tets} shows that the that the disc lifetime is shortened from $\sim$100kG. We now use numerical orbit calculations to examine the circularisation timescale and lifetime of eccentric discs.

\subsection{Results of Simulations}\label{sec:ROS} 

We ran a set of simulations with the same initial conditions as \cite{veras_leinhardt_2014} from section \ref{sec:4_nbodycode}, which gave a pericentre distance (or impact factor) as $\approx$15 WD radii or 0.15$R_{\odot}$. The accretion radius was kept at 2 WD radii. We found, at these close approaches, the influence of the field was too strong to provide meaningful results. So in addition, we ran a second set of simulations where the apocentre (0.4AU) remained the same, but adjusted the eccentricity ($e=0.9884$) so the pericentre distance was at 50 WD radii (0.5$R_\odot$). \medskip

\begin{landscape}
\begin{figure}
    \centering
    \includegraphics[width=\linewidth]{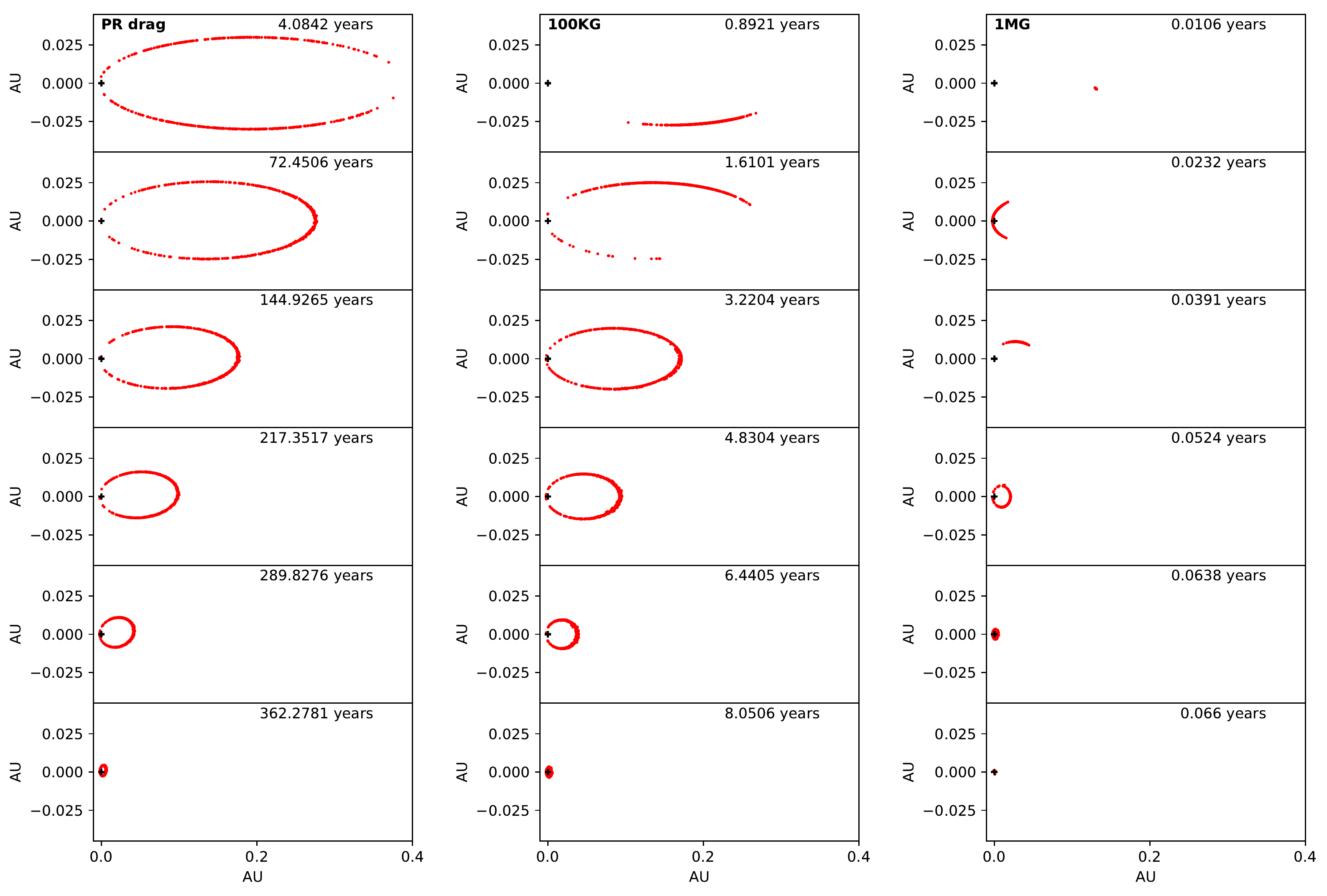}
    \caption{This plot shows results from three simulations; PR drag alone, 100KG and 1MG. It shows the lifetime is decreased as a result of the magnetic drag and faster eccentricity damping compared to the PR case.}
    \label{fig:big_plot}
\end{figure}
\end{landscape}

\begin{table}
\begin{tabular}{lll}
\textbf{B Field Strength} & \multicolumn{2}{l}{\textbf{Disc Lifetime (Years)}} \\ \hline 
                                 \multicolumn{1}{r}{$R_{peri}$ =} & 0.15$R_{\odot}$ & 0.5$R_{\odot}$  \\ \hline
100 MG & $5.3\times10^{-4}$ & $1.14\times10^{-3}$\\
10 MG & $5.0\times10^{-4}$ & $1.17\times10^{-3}$\\
1 MG & $5.0\times10^{-4}$ & $6.61\times10^{-2}$\\
100 KG & $1.5\times10^{-3}$ & 8.247 \\
10 KG & 0.245 & 248.826 \\
1 KG & 16.52 & 379.563  \\
PR & 41.94 & 379.563 
\end{tabular}
\caption{This table gives the lifetimes of discs at different field strengths with two different pericentre distances, with an apocentre of 0.4AU}
\label{tab:results}
\end{table}


\subsubsection{PR Drag}
We ran a simulation with unshielded PR drag alone to find the maximum amount of time a body composed of micron sized dust can survive in the 15 and 50 $R_{WD}$ cases. The lifetime of the 15 $R_{WD}$ is $\approx$40 years and increases to $\approx$380 years for the 50 $R_{WD}$. There is a 3.3 factor pericentre increase in distance between the 15 and 50 $R_{WD}$ but a 9.5 factor increase in lifetime. In both cases the particles remain in a narrow annulus for the majority of their lifetime, then circularise and accrete rapidly in the last few years. The PR drag acts to first damp the eccentricity and circularise the disc inside the tidal radius before accretion begins.

\subsubsection{1KG}
The 1KG case is interesting because at the higher impact factor of 15$R_{WD}$ the lifetime is $\approx$16 years, 2.5 times shorter than the PR drag case. This was expected from Figure \ref{fig:Tlife_tets}. Similarly, the 50 impact factor has the same lifetime as the PR drag as expected; $\approx$380 years. Fields strengths 1kG and lower therefore play a small role in debris disc formation.  

\subsubsection{10KG}
These simulations produce a similar result to 1kG runs.A pericnetre of 15$R_{WD}$ gives a lifetime of $\approx$0.2 years and the 50 $R_{WD}$ is $\approx$250. This confirms the models prediction that for highly eccentric orbits, 10kG should begin to lower the circularisation and disc lifetime.

\subsubsection{100KG}
The 100KG has short lifetimes in both the 15 $R_{WD}$ case, $\approx$0.0015 years, and 50 $R_{WD}$ case, $\approx$8 years. The strength of the magnetic field completely dominates the lifetime of the disc at both distances and the drag is strong enough that the particles no longer initially trace the original orbit. This is expected as we found in equation \ref{eq:Nevfla}
that the point of magnetic dominance occurs at 100KG for a disc at 1$R_{\odot}$, any field strength above this will be completely magnetically dominated for this temperature. Figure \ref{fig:big_plot} shows a comparison between PR drag, 100kG and 1MG to illustrate the enhanced eccentricity damping and shortened lifetime due to the magnetic drag.

\subsubsection{1MG}
The 1MG field is strong enough that in both impact factor cases the particles are only able to orbit the white dwarf a handful of times before they are accreted. The magnetic field is strong enough that the particles are noticeably decelerated during the pericentre passes so that the orbit is significantly compressed inwards each orbit.

\subsubsection{10MG and 100MG}
The 10MG and 100MG cases both give similarly short lifetimes. The magnetic field is strong enough that the particles experience extreme deceleration in the first close approach and accrete onto the white dwarf immediately. No disc forms in these regimes due to the extreme forces on the micron sized particles. For these two regimes we also did a third set of simulations where the pericentre was just inside the tidal radius and found that at this distance the two regimes do separate in lifetime. The 100MG case survives $\approx$0.003 years and the 10MG case gives a lifetime of $\approx$0.014 years. The lifetime of the 100MG case is still short enough that no disc will form as the particles survive just a handful of orbits. In the 10MG case the particles survive 8 orbits before they are accreted, however each orbit is strongly damped and circularises before accretion occurs.

\begin{figure*}
    \centering
    \includegraphics[width=\linewidth]{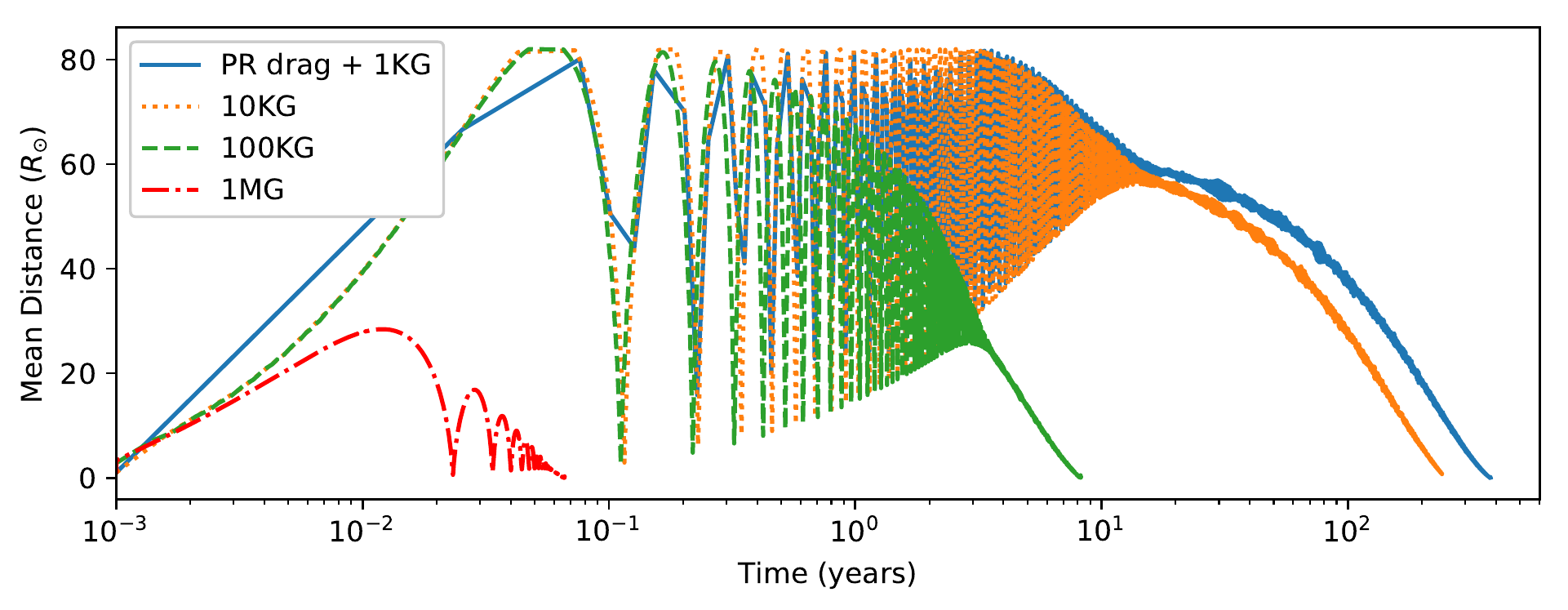}
    \caption{Evolution of the mean particle distance for PR drag, 1KG, 10KG, 100KG, and 1MG. With an initial pericentre of 50 $R_{WD}$.}
    \label{fig:x_evolution}
\end{figure*}

\begin{figure*}
    \centering
    \includegraphics[width = \linewidth]{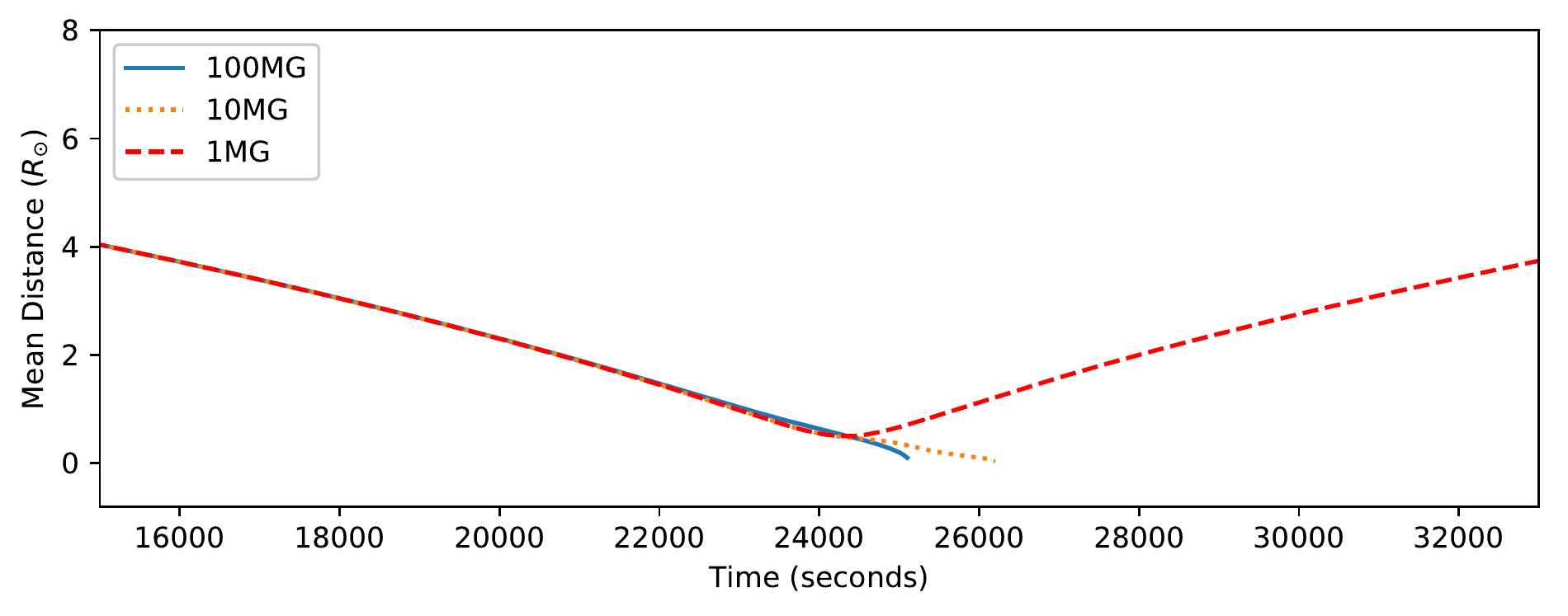}
    \caption{Evolution of the mean particle distance for 1MG, 10MG, 100MG. With an initial pericentre of 50 $R_{WD}$.}
    \label{fig:MG_evolution}
\end{figure*}

\subsection{Disc Lifetime}
We directly compare the effect of the differing field strengths and PR drag by comparing the particle loss over time (See figures \ref{fig:x_evolution} and \ref{fig:MG_evolution}). We consider the disc to be depleted when there is less than 10\% of the original material left. 
The 100KG and MG cases all show magnetically dominated schemes at all initial pericentre distances.
In the PR drag, 1KG, and 10KG cases, we see that they PR drag has some effect depending on the distance of the pericentre. Specifically we can see that for this particular PR drag (i.e 10\,000K) that 10KG is the point where the dominant force changes between the two. 
This roughly matches the analytical results from Figure \ref{fig:Tlife_tets} where the PR drag should dominate below approximately 25KG for a 10,000K white dwarf when the pericentre distance is just inside the tidal radius. We show in our simulations that 10KG is where the dominate force begins to change for close in pericentre distances of 15 and 50$R_{WD}$. We can therefore approximate that the change in dominant force occurs between 10-25KG, for micron sized particles, depending on the impact factor of the particles.

In Figures \ref{fig:x_evolution} and \ref{fig:MG_evolution} we see the decay rate of the particles x value over time. The graph illustrates the average displacement along the x axis of all the particles for kG and MG regimes. At the beginning of the low field strength simulations (Fig \ref{fig:x_evolution}) we see that the average x value oscillates as the particles are fairly close together and haven't spread out over the orbit. This means, until the particles have spread out, the average particle displacement will trace the orbit. Once the orbit is filled we see a roughly linear decline in average x distance over time as the disc is compressed inwards and the eccentricity of the disc is damped. The figure shows that the average x displacement in the simulations will decay quicker with a high magnetic field, while the PR dominated and low magnetic field cases are relatively slow. In Figure \ref{fig:MG_evolution} we see the rapidity of particle accretion for the MG cases, the 10 and 100MG cases are only distinguishable by a few thousand seconds at 50$R_{WD}$ as they accrete during the first pericentre pass.

\subsection{Pericentre dependence}
\begin{figure}
    \centering
    \includegraphics[width=\linewidth]{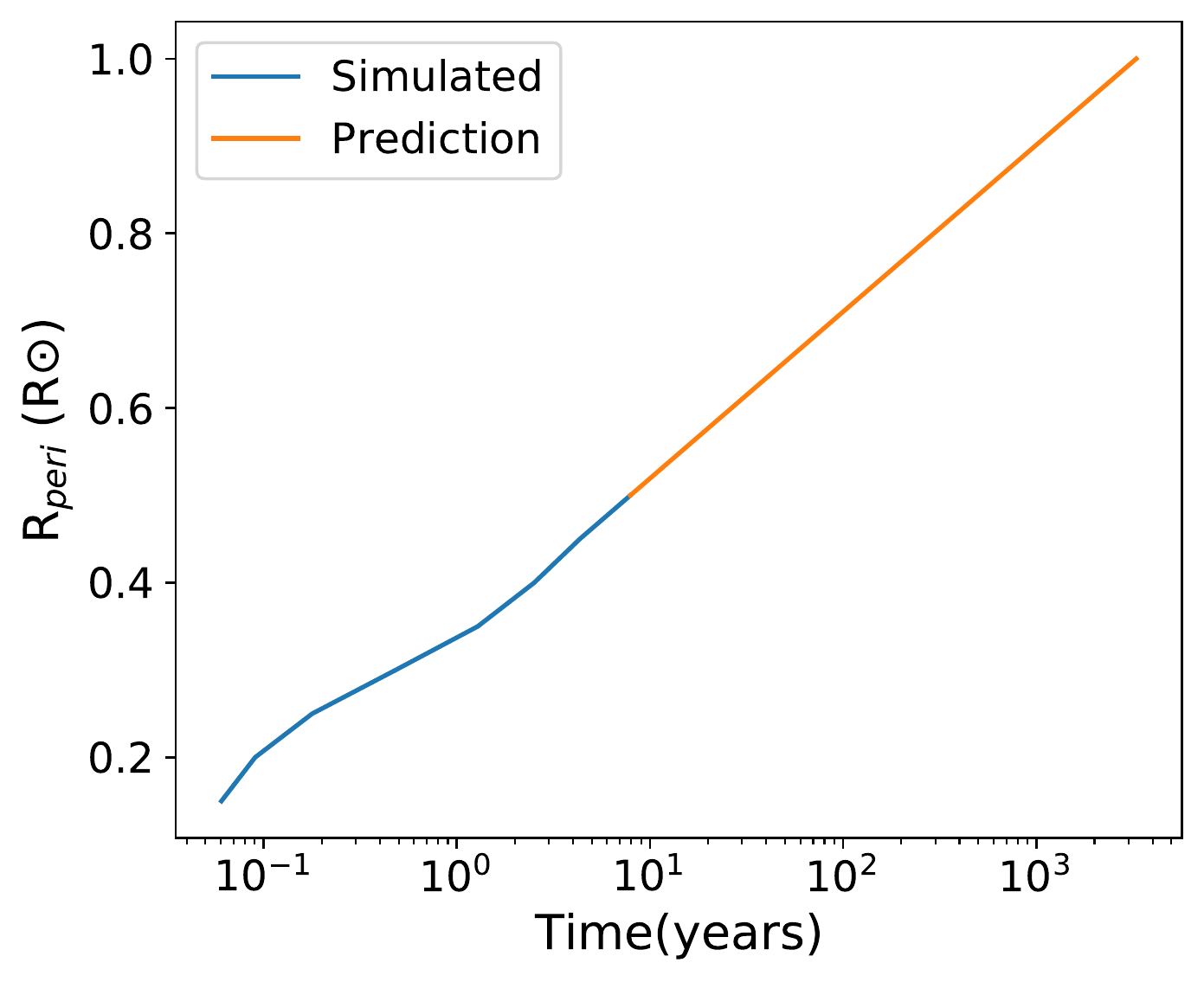}
    \caption{Particle Lifetime as a function of pericentre around a 100kG WD and apocentre of 0.4AU}
    \label{fig:peri_prediction}
\end{figure}

We ran a set of single particle simulations at 100KG with unshielded PR drag and a 0.4AU apocentre. We adjusted the eccentricity to sweep from 15-50$R_{WD}$ as a pericentre distance to study how the pericentre distance effected the particle lifetime. We found an exponential relation which we extrapolated out to 100$R_{WD}$ (shown in Figure \ref{fig:peri_prediction}). Our results show that the lifetime of the particle varies between $\sim 10^{-3}$ years up to $10^3$ years depending on the pericentre distance.

\section{Discussion} \label{sec:DISS}

Looking at the simulations, Figure \ref{fig:big_plot} shows PR drags matches the expected behaviour predicted by \citet{veras_leinhardt_2015}; a narrow eccentric ring of debris, where the eccentricity is damped until the disc is circularised and then accretion occurs. The magnetic drag model maintains this behaviour, the difference being the magnitude of the damping is more severe, emphasised in Figure \ref{fig:x_evolution}. This result is supported by the observations of \cite{vanderbosch_hermes_2019}, where the dust of a disrupted asteroid is following an eccentric orbit. For fields $>$10MG the drag force is strong enough to preventing the formation of a discernible disc (Figure \ref{fig:MG_evolution}).

For eccentric orbits, there is a strong dependence for disc lifetime on the pericentre. Figure \ref{fig:peri_prediction} shows the relationship between disc lifetime as a function of pericentre distance based on our simulations. We extrapolate this relationship for wider, more typical, pericentre distances. Dividing the mass of a Ceres sized asteroid  ($10^{22}$ g)  by these lifetimes, we can find a crude accretion rate of $10^{18}$ g s$^{-1}$ at $15 R_{WD}$ down to $10^{7.5}$ g s$^{-1}$ at $100 R_{WD}$ for micron sized particles. These accretion rates would decrease for larger particle sizes. This indicates that disc lifetimes would be highly dependant on magnetic field strength, pericentre distance, and particle size. Discs with a moderate magnetic field of 100kG can easily have similar accretion rates to Figure \ref{'fig:PR_plot'} if the disc apocentre is close to the tidal radius or is made up of larger particles.

\begin{figure}
    \centering
    \includegraphics[width=\linewidth]{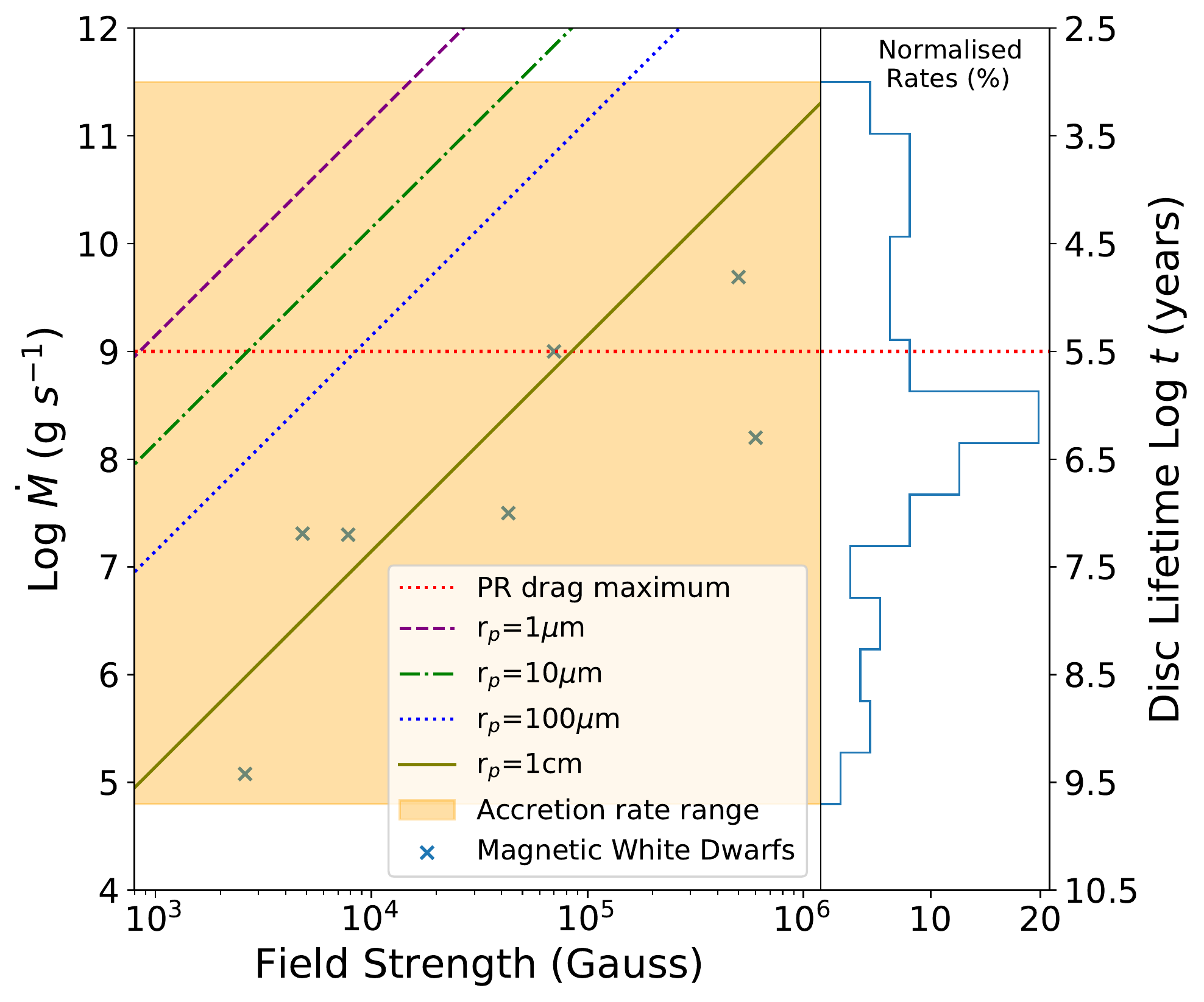}
    \caption{Predicted accretion rates with diamagnetic drag model from the analytical equations}
    \label{fig:pred_rates}
\end{figure}

Using the same logic for Figure \ref{fig:magfield_disclife}, we can estimate the accretion rates for different sized particles in a circular orbit at 1R$_\odot$. Figure \ref{fig:pred_rates} gives the predicted accretion rates at different field strengths and the corresponding disc lifetimes. The plot also shows observed magnetic WDs and a histogram of accretions rates from Figure \ref{'fig:PR_plot'}. The predicted rates for centimetre sized particles are in closest agreement with the observed accretion rates. It is perhaps interesting to note that while discs are observed to have many micron sized particles, the majority of the mass may reside in larger particles that drive the accretion rate.

We employed a Spearman's rank test on the magnetic systems and found a positive correlation (R$_s$ = 0.88, n = 7, p < 0.05) between the field strength and accretion rate.

\begin{figure}
    \centering
    \includegraphics[width=\linewidth]{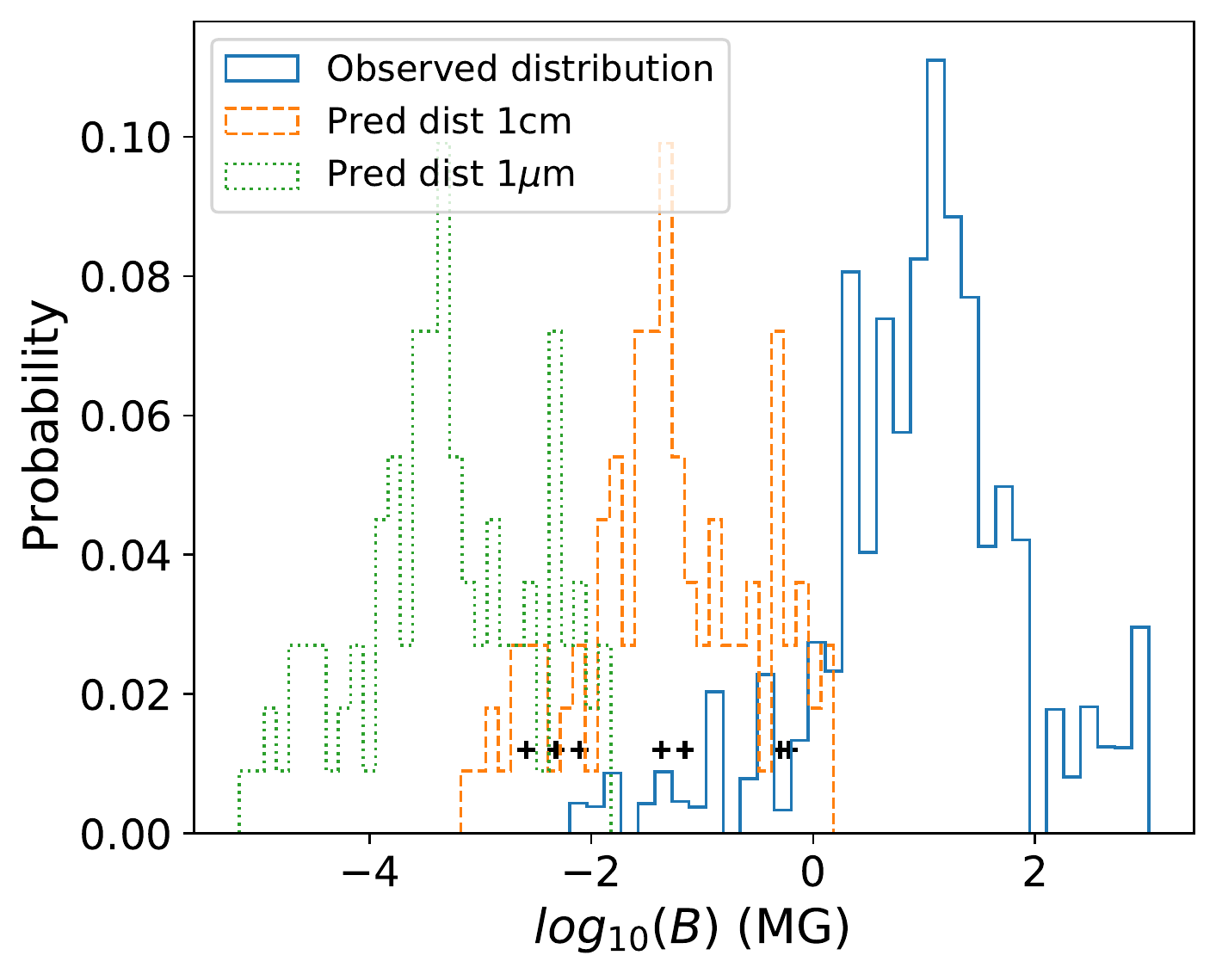}
    \caption{Predicted distribution of field strengths in WDs based on the inferred accretion rates from figure \ref{'fig:PR_plot'} and the observed distribution of magnetic WDs from figure \ref{fig:distribution}. The crosses indicate the field strengths of observed WDs with associated accretion rates.}
    \label{fig:pred_mag_dist}
\end{figure}

Given that 2-20\% of WDs are observed to be magnetic and up to a third of WDs are polluted, we expect $\sim$0.6-7\% of WDs to be both polluted and magnetic, and therefore have noticeably shorter disc lifetimes. This is assuming no causal relation between magnetism and pollution in WDs. While this does not account entirely for the lack of observed discs, it may suggest that magnetic fields form a piece of the missing disc puzzle. We see from Figure \ref{fig:pred_rates} that reasonable accretion rates can be achieved well below the detectable threshold for magnetic fields. The rates predicted for WDs with non-detectable field strengths still compensate where shielded PR-drag falls short. If we assume these accreting systems are driven by magnetism, we can convolve the predicted accretion rates from Figure \ref{fig:pred_rates} with the distribution of inferred accretion rates to create a distribution of field strengths in accreting WDs. The results of this are presented in Figure \ref{fig:pred_mag_dist}. If these systems truly are magnetic, they are most likely in the kG regime, which is close to or beneath the detection limit. Comparing the observed field strengths to the predicted distribution for centimetre sized particles in Figure \ref{fig:pred_mag_dist}, there is an overlap between the two. This could indicate there is a population of accreting WDs with magnetic fields that have gone undetected because they are below the detection threshold. One interpretation of this is that most WDs exhibit some level of magnetism, but this magnetism is unresolvable with current methods. It is important to note here that the true distribution of magnetic WDs is not known as the observations suffer from a lack of completeness, especially for weaker field strengths.

As mentioned earlier, we observe around one third of WDs as polluted and between 0.6-6\% of those have a disc. Given that sinking timescales are much shorter than expected disc lifetimes, we should see a much higher rate of polluted WDs with an accompanying disc; the missing disc problem. We will now investigate the impact diamagnteic drag has on the missing disc problem. Given that a third of WDs are polluted we define probability of pollution as the observed lifetime of the disc, $t_{disc}$, and sinking timescale, $t_{sink}$, divided by the time between impacts, $t_{imp}$:
\begin{equation}
    P_{(pol)} = \frac{t_{disc}+t_{sink}}{t_{imp}} \approx \frac{1}{3}.
\end{equation}

Rearranging the above, we get an expression for the time between impact events, $t_{imp}$:

\begin{equation}
    t_{imp} = 3\left(t_{disc}+t_{sink}\right).
\end{equation}

We can then derive a probability of pollution with no visible discs as:
\begin{equation}
    P_{(pol,no\:disc)} = \frac{t_{sink}}{t_{imp}} \sim \frac{1}{3} \frac{t_{sink}}{t_{disc}+ t_{sink}}.
    \label{EQ:P_DIDDY}
\end{equation}

\begin{figure}
    \centering
    \includegraphics[width=\linewidth]{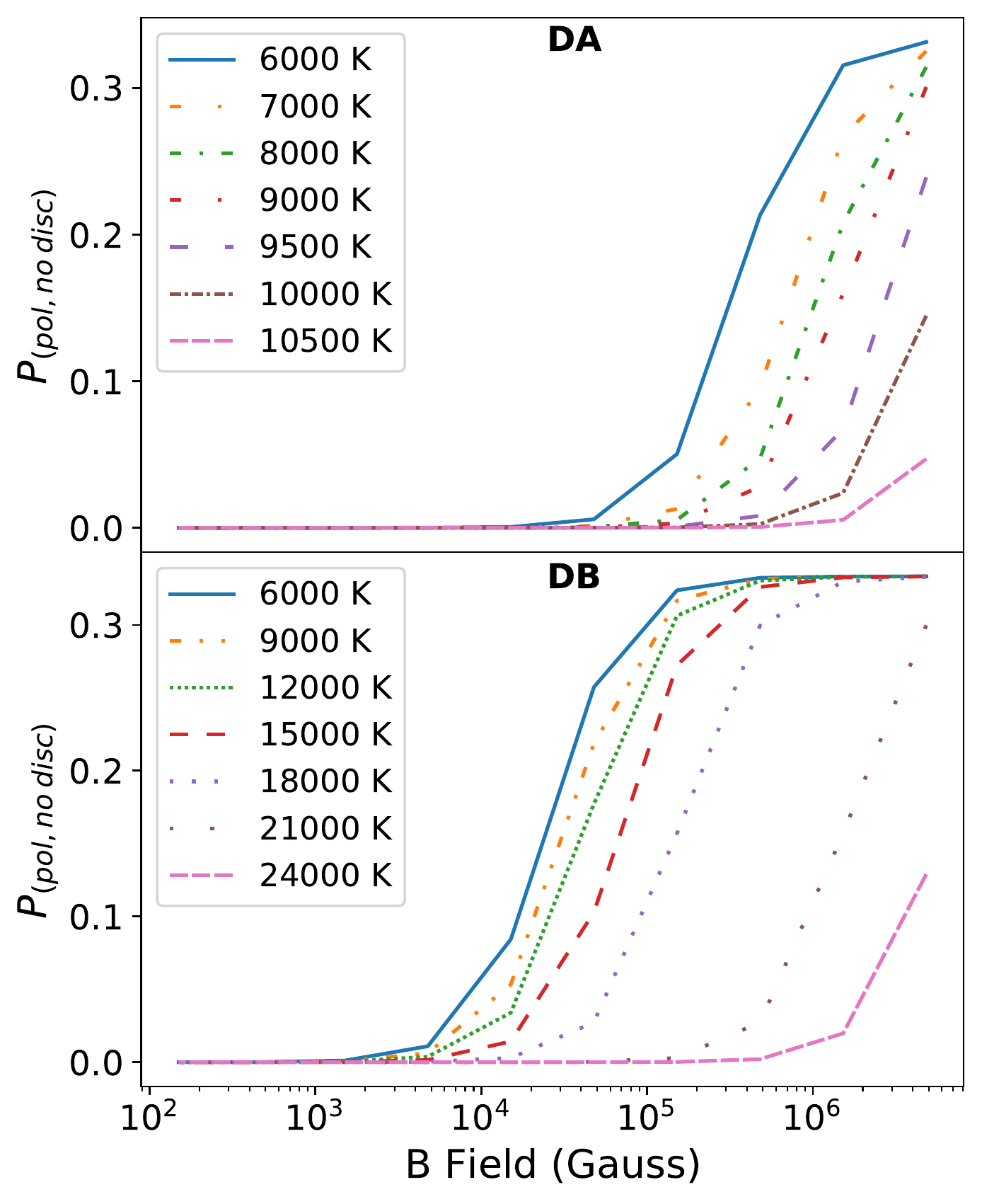}
    \caption{The probability of observing a WD that is both polluted and without a disc based on the relationship between disc lifetime for 1cm particles and magnetic field strength for different sinking timescales in both DAs and DBs.}
    \label{fig:DA_DB_STATS}
\end{figure}

We find the probability of a WD being polluted without a disc for different lifetimes using \ref{EQ:P_DIDDY}. Using the analytical relationship between disc lifetime and magnetic field strength (equation \ref{eq: Life}) for centimetre sized particles, we find the probability of seeing a polluted WD without a disc for different field strengths over a range of sinking timescales. Figure \ref{fig:DA_DB_STATS} shows the probability of a WD being polluted with no observable disc as a function of magnetic field.

The upper plot shows that for DA WDs, the magnetic field would need to be $>10^5$ Gauss to lower the disc lifetime enough that pollution could be seen without the presence of a disc. This is unlikely as Mega-Gauss strength fields are easily observable. The bottom plot shows the same data but for DB WDs which decreases the required magnetic field strength to $>10^4$ Gauss. For DB type stars, the lack of observable disc could be explained with low magnetic fields that are difficult to observe with current techniques. This shows that magnetic fields may explain some of the missing discs, but cannot explain the majority as most WDs would need measurably high field strengths. 

It is suspected that convection in the atmosphere is likely to be slowed or stopped by a magnetic field and could potentially cause pollution to be visible in the atmosphere for longer than gravitation settling estimates \citep{tremblay2015evolution, gentile2018can, ferrario_wickramasinghe_2020}. The outcome of both magnetic drag and convective damping together is that the disc would be shorter lived and increase the accretion rate, while the convective damping would make the pollution last longer after the disc has been consumed which could decrease the historical or inferred accretion rates.

Variability in observed discs may also be explained by magnetism \citep{swan_fraihi_2019}. Smaller particles are effected more strongly than larger ones in both PR drag and magnetic fields and with a fast rotating field the different size particles may get trapped in the magnetosphere, leading to variability. Similarly, if the dipole field is tilted the amount of the particles in view changes over the orbit, potentially causing variable light curves. This work only included a stationary magnetic field which exerted a drag force in all circumstances. However, observations of WD 1145+017 indicate signposts of dust trapping, which can only be caused by co-rotation \citep{farihi_vonHippel_2017}. Future work can simulate different spins to test this trapping model as well as the reaction of particles to a rapidly rotating WD. \\

\subsection{Limitations of the Model}
The simulations in this work uses micron sized particles so the effect of the magnetic field is relatively strong, realistically the asteroid would break up into chunks of varying sizes which would slowly get ground into smaller pieces before reaching sizes where the magnetic field would have an effect. By using micron sized particles from the beginning, the stage of the disruption and formation of the disc is sped up significantly. Using the same reasoning, PR drag is strong for particles smaller than a centimetre; so the PR drag is particularly effective on the micron sized particles used. It should be noted that as the particles are all the same size and we have not included collisions, thus the particles will accrete on similar timescales. Using an array of different sized particles would likely result in particles accreting at different times and show signs of disc filling as opposed to an annulus seen in these simulations. 

We use a small orbit for these simulations with a semi-major axis of 0.2AU, we expected that larger, more realistic orbits would create longer lived discs in all cases as the particles will spend more time outside of the magnetic field's influence. However, we expect that the relation seen in the simulations would hold for larger orbits as time spent in the tidal radius is nearly independant of semi-major axis \citep{veras_leinhardt_2014}.

Shielded PR drag occurs where particles in the disc shield other particles from the full force of the stellar radiation, which acts to slow the accretion and extend disc lifetime \citep{rafikov_2011a, farihi_2016}. The magnetic drag is not effected in this way as all particles are affected based on their distance from the field and do not shield one another. Compared to the unshielded PR drag we have used in this work we would expect shielded PR drag to be an even larger contrast to magnetic drag in terms of disc lifetime and accretion rate. Due to the time needed to simulate shielded PR drag and the small number of particles used we decided to compare to unshielded PR drag as the lifetimes involved are shorter. However, we would expect the disc lifetimes to be shortened by the magnetic drag down to field strengths of a few kG.   

We do not include the effects of collisions in this study. Through preparatory simulations of 50\,000 particles in an eccentric configuration, we found that the number of collisions were on the order of 10 over 5 years of simulation time. This is most likely due to small particles being on large orbits, making the impact of collisions negligible. This is echoed in \cite{veras_leinhardt_2014}. When addressing the more realistic tidal disruption scenario of non-uniform particles and the influence of a tilted magnetic field, collisions may become more important due to the increased particle density and chaotic particle pathways. 

We also do not incorporate sublimation. As dust gets closer to the WD it will sublimate into gas (at $\sim$1\,900K), which in turn becomes ionised and follows the magnetic field lines. We estimate that carbonaceous and silicate dust would sublimate at a distance between 9-13 WD radii (for a 10,000K WD and depending on absorption) and thus, our grains would not sublimate at the 15 and 50$R_{WD}$ pericentre distances we use in our simulations until the disc circularises and starts compressing inwards \citep{kobayashi_2011}. The effect of sublimation on the disc lifetime in this case is therefore minimal compared to the effect of the magnetic field and PR drag. However, for warmer WDs where heating and sublimation could take place further from the central WD, disc lifetimes would be further decreased. For simplicity, we do not include sublimation in this work as we focused primarily on the dust dynamics. Though a detailed study on accretion dynamics would require the inclusion of dust-gas interaction and sublimation. 

We use a WD that has a 0.5 Gyrs cooling age and has cooled to $\sim$ 10\,000K, therefore the luminosity is quite low compared to a younger WD. The PR drag in this case is rather weak and would get weaker for the older, cooler WDs. Here we find that the $<$10KG magnetic field strengths are weak compared to the PR drag, but it is possible that in an older WD weaker magnetic fields can dominate over the PR drag. In the same vein, a younger WD will have a stronger PR drag effect which could be more dominant than the higher magnetic field strengths. Studies by \citet{kawka_vennes_2014} and \cite{ferrario_wickramasinghe_2020} find a high incidence of magnetism in polluted, cooler WDs where it is likely that the low KG regimes will dominate.

\subsubsection{Comments on the Simulations}
Tests were conducted on how the number of particles effected the simulations. The analytical results had no dependence on particle number so the lifetime of the disc should be the same, regardless of the particle number. Indeed, as shown in Appendix Figure \ref{fig:diff_part_num}, we find that the lifetimes are identical for 1, 5, 10, 50, 100, 500, 1000, 5000 and 10\,000 particles. The only change is the time it takes for the simulation itself to run. We can use this to speed up the simulations without effecting our results in the lower magnetic field simulations which have longer disc lifetimes. This exploitation is only viable with test particles (particles that do not interact), when particle-particle interactions are included the number of particles will effect disc structure and lifetime in the simulations. 

Furthermore, we tested the best time step to use for our simulations. We tested: 0.1, 0.5, 1, 5, 10, 15, 20, 25, 30, and 50 second time steps for the 1MG field strength and found no difference up to 10 seconds, beyond that numerical error overwhelms the simulation and ejects the particles (see Appendix Fig \ref{fig:diff_time_step}). We can conclude that the numerical errors for a step size up to 2.5 seconds are negligible for the simulations used for this project; however, 10 seconds can be used for any field strengths less than 1MG.

\section{Conclusions}\label{sec:Conclusion}
In this paper, we have highlighted that moderate strength magnetic fields may play an important role in determining the dynamics of the circularisation and accretion of a debris disc onto a WD. Using an analytic model and numerical orbit calculations we were able to simulate the disruption and subsequent accretion of a disrupted asteroid. We estimated the disc lifetimes and accretion rates driven by the magnetic drag force and showed field can reduce the disc lifetime significantly for field strengths typically observed in WDs. 
We found the following main results from this work:
\begin{itemize}
    \item Disc lifetime is decreased by magnetic fields above $\sim$10KG for optically thick discs and $\sim$100kG for thin discs.
    \item The lifetime of the particles has a strong dependence on the pericentre distance, especially in the kG regimes.
    \item In both the PR drag and magnetic dominanted scenarios the eccentricity is damped and the disc is circularised before accretion occurs, except for the very highest magnetic field strengths.
    \item The addition of a magnetic field increases the eccentricity damping and therefore decreases the circularisation timescale.
    \item The particles used in our simulations are micron sized which are very strongly effected by both magnetic fields and PR drag. Particle size is important in deciding the magnitude of the force on the particle, but not which term, PR or magnetic, is dominant. 
    \item The shortened lifetimes and addition of magnetic drag leads to increased accretion rates in magnetic WDs.
    \item Using the observed accretion rates, we can infer a distribution of magnetic field strengths in WDs. Tentatively, we can see there is a possibility that many polluted WDs may exhibit some magnetism; however, most of these are below the detectable limit.
    \item The diamagnetic drag force cannot solve the missing disc problem, although it may account for some DB and cool DA stars with missing discs. 
    \item There is a weak positive correlation in accretion rate and magnetic field strength in observed magnetic polluted white dwarfs.
\end{itemize}

\section*{Acknowledgements}
Special thanks to our anonymous reviewer and Dimitri Veras for helpful comments to improve the manuscript. We thank Matthew Hoskin for insights into historical accretion, and Roman Rafikov for his explanation surrounding shielded PR drag. This research was made possible thanks to funding from the Science and Technologies Facilities Council (STFC).
This work was performed using the Cambridge Service for Data Driven Discovery (CSD3), part of which is operated by the University of Cambridge Research Computing on behalf of the STFC DiRAC HPC Facility (www.dirac.ac.uk). The DiRAC component of CSD3 was funded by BEIS capital funding via STFC capital grants ST/P002307/1 and ST/R002452/1 and STFC operations grant ST/R00689X/1. DiRAC is part of the National e-Infrastructure.
This research used the ALICE High Performance Computing Facility at the University of Leicester.

\section{Data Availability}
These results were calculated using the code \texttt{Freefall}, available on \href{https://github.com/Ry-C123/Freefall}{Github} 

\bibliographystyle{apalike}
\bibliography{MahPapers}

\begin{thebibliography}{}

\bibitem[{Ablimit} and {Maeda}, 2019]{ablimit_maeda_2019}
{Ablimit}, I. and {Maeda}, K. (2019).
\newblock {Evolution of Magnetized White Dwarf Binaries to Type Ia Supernovae}.
\newblock {\em \apj}, 871(1):31.

\bibitem[{Alcock} et~al., 1986]{alcock_fristrom_1986}
{Alcock}, C., {Fristrom}, C.~C., and {Siegelman}, R. (1986).
\newblock {On the number of comets around other single stars}.
\newblock {\em \apj}, 302:462--476.

\bibitem[{Alecian} et~al., 2007]{alecian_wade_2007}
{Alecian}, E., {Wade}, G.~A., {Catala}, C., {Folsom}, C., {Grunhut}, J.,
  {Donati}, J.~F., {Petit}, P., {Bagnulo}, S., {Boehm}, T., {Bouret}, J.~C.,
  and {Landstreet}, J.~D. (2007).
\newblock {Magnetism, rotation and accretion in Herbig Ae-Be stars}.
\newblock In {Bouvier}, J. and {Appenzeller}, I., editors, {\em Star-Disk
  Interaction in Young Stars}, volume 243 of {\em IAU Symposium}, pages 43--50.

\bibitem[Bardeen et~al., 1957]{bardeen1957}
Bardeen, J., Cooper, L.~N., and Schrieffer, J.~R. (1957).
\newblock Theory of superconductivity.
\newblock {\em Physical review}, 108(5):1175.

\bibitem[Barstow et~al., 1995]{barstow1995}
Barstow, M., Jordan, S., O’Donoghue, D., Burleigh, M., Napiwotzki, R., and
  Harrop-Allin, M. (1995).
\newblock Re j0317--853: the hottest known highly magnetic da white dwarf.
\newblock {\em Monthly Notices of the Royal Astronomical Society},
  277(3):971--985.

\bibitem[{Bauer} and {Bildsten}, 2018]{bauer_bildsten_2018}
{Bauer}, E.~B. and {Bildsten}, L. (2018).
\newblock {Increases to Inferred Rates of Planetesimal Accretion due to
  Thermohaline Mixing in Metal-accreting White Dwarfs}.
\newblock {\em \apjl}, 859:L19.

\bibitem[Berry and Geim, 1997]{berry1997}
Berry, M.~V. and Geim, A.~K. (1997).
\newblock Of flying frogs and levitrons.
\newblock {\em European Journal of Physics}, 18(4):307.

\bibitem[{Bochkarev} and {Rafikov}, 2011]{bochkarev_rafikov_2011}
{Bochkarev}, K.~V. and {Rafikov}, R.~R. (2011).
\newblock {Global Modeling of Radiatively Driven Accretion of Metals from
  Compact Debris Disks onto White Dwarfs}.
\newblock {\em \apj}, 741:36.

\bibitem[{Bonsor} et~al., 2017]{bonsor_farihi_2017}
{Bonsor}, A., {Farihi}, J., {Wyatt}, M.~C., and {van Lieshout}, R. (2017).
\newblock {Infrared observations of white dwarfs and the implications for the
  accretion of dusty planetary material}.
\newblock {\em \mnras}, 468:154--164.

\bibitem[{Bonsor} and {Veras}, 2015]{bonsor_veras_2015}
{Bonsor}, A. and {Veras}, D. (2015).
\newblock {A wide binary trigger for white dwarf pollution}.
\newblock {\em \mnras}, 454:53--63.

\bibitem[{Bouvier} et~al., 2007]{bouvier_alencar_2007}
{Bouvier}, J., {Alencar}, S.~H.~P., {Harries}, T.~J., {Johns-Krull}, C.~M., and
  {Romanova}, M.~M. (2007).
\newblock {Magnetospheric Accretion in Classical T Tauri Stars}.
\newblock In {Reipurth}, B., {Jewitt}, D., and {Keil}, K., editors, {\em
  Protostars and Planets V}, page 479.

\bibitem[{Briggs} et~al., 2018]{briggs_ferrario_2018}
{Briggs}, G.~P., {Ferrario}, L., {Tout}, C.~A., and {Wickramasinghe}, D.~T.
  (2018).
\newblock {Genesis of magnetic fields in isolated white dwarfs}.
\newblock {\em \mnras}, 478(1):899--905.

\bibitem[{Briggs} et~al., 2015]{briggs_ferrario_2015}
{Briggs}, G.~P., {Ferrario}, L., {Tout}, C.~A., {Wickramasinghe}, D.~T., and
  {Hurley}, J.~R. (2015).
\newblock {Merging binary stars and the magnetic white dwarfs}.
\newblock {\em \mnras}, 447(2):1713--1723.

\bibitem[{Bromley} and {Kenyon}, 2019]{bromley_kenyon_2019}
{Bromley}, B.~C. and {Kenyon}, S.~J. (2019).
\newblock {Ohmic Heating of Asteroids around Magnetic Stars}.
\newblock {\em \apj}, 876(1):17.

\bibitem[{Burns} et~al., 1979]{burns_lamy_1979}
{Burns}, J.~A., {Lamy}, P.~L., and {Soter}, S. (1979).
\newblock {Radiation forces on small particles in the solar system}.
\newblock {\em \icarus}, 40:1--48.

\bibitem[{Carry}, 2012]{carry_2012}
{Carry}, B. (2012).
\newblock {Density of asteroids}.
\newblock {\em \planss}, 73(1):98--118.

\bibitem[{Chancia} et~al., 2019]{chancia_hedman_2019}
{Chancia}, R.~O., {Hedman}, M.~M., {Cowley}, S.~W.~H., {Provan}, G., and {Ye},
  S.~Y. (2019).
\newblock {Seasonal structures in Saturn's dusty Roche Division correspond to
  periodicities of the planet's magnetosphere}.
\newblock {\em \icarus}, 330:230--255.

\bibitem[{Chandrasekhar}, 1931]{chandrasekher_1931}
{Chandrasekhar}, S. (1931).
\newblock {The Maximum Mass of Ideal White Dwarfs}.
\newblock {\em \apj}, 74:81.

\bibitem[{Collier Cameron} and {Campbell}, 1993]{collier_campbell_1993}
{Collier Cameron}, A. and {Campbell}, C.~G. (1993).
\newblock {Rotational evolution of magnetic T Tauri stars with accretion
  discs}.
\newblock {\em \aap}, 274:309.

\bibitem[{Cropper}, 1990]{cropper_1990}
{Cropper}, M. (1990).
\newblock {The Polars}.
\newblock {\em \ssr}, 54(3-4):195--295.

\bibitem[Cunningham et~al., 2019]{cunningham2019}
Cunningham, T., Tremblay, P.-E., Freytag, B., Ludwig, H.-G., and Koester, D.
  (2019).
\newblock Convective overshoot and macroscopic diffusion in
  pure-hydrogen-atmosphere white dwarfs.
\newblock {\em Monthly Notices of the Royal Astronomical Society},
  488(2):2503--2522.

\bibitem[Cutter and Hogg, 2020]{cutter_Hogg2020}
Cutter, R. and Hogg, M. (2020).
\newblock {Freefall\_Beta}.
\newblock {\em {Zenodo}}.

\bibitem[{Deal} et~al., 2013]{deal_deheuvels_2013}
{Deal}, M., {Deheuvels}, S., {Vauclair}, G., {Vauclair}, S., and {Wachlin},
  F.~C. (2013).
\newblock {Accretion from debris disks onto white dwarfs. Fingering
  (thermohaline) instability and derived accretion rates}.
\newblock {\em \aap}, 557:L12.

\bibitem[Debes and Sigurdsson, 2002]{debes_sigurdsson_2002}
Debes, J.~H. and Sigurdsson, S. (2002).
\newblock Are there unstable planetary systems around white dwarfs?
\newblock {\em The Astrophysical Journal}, 572(1):556.

\bibitem[Debes et~al., 2007]{debes2007}
Debes, J.~H., Sigurdsson, S., and Hansen, B. (2007).
\newblock Cool customers in the stellar graveyard. iv. spitzer search for
  mid-ir excesses around five das.
\newblock {\em The Astronomical Journal}, 134(4):1662.

\bibitem[{Desharnais} et~al., 2008]{desharnais_wesemael_2008}
{Desharnais}, S., {Wesemael}, F., {Chayer}, P., {Kruk}, J.~W., and {Saffer},
  R.~A. (2008).
\newblock {FUSE Observations of Heavy Elements in the Photospheres of Cool DB
  White Dwarfs}.
\newblock {\em \apj}, 672:540--552.

\bibitem[Drell et~al., 1965]{drell_foley_1965}
Drell, S.~D., Foley, H.~M., and Ruderman, M.~A. (1965).
\newblock Drag and propulsion of large satellites in the ionosphere; an
  alfv\'en propulsion engine in space.
\newblock {\em Phys. Rev. Lett.}, 14:171--175.

\bibitem[{Dufour} et~al., 2007]{dufour_bergeron_2007}
{Dufour}, P., {Bergeron}, P., {Liebert}, J., {Harris}, H.~C., {Knapp}, G.~R.,
  {Anderson}, S.~F., {Hall}, P.~B., {Strauss}, M.~A., {Collinge}, M.~J., and
  {Edwards}, M.~C. (2007).
\newblock {On the Spectral Evolution of Cool, Helium-Atmosphere White Dwarfs:
  Detailed Spectroscopic and Photometric Analysis of DZ Stars}.
\newblock {\em \apj}, 663:1291--1308.

\bibitem[{Dupuis} et~al., 1992]{dupuis_fontaine_1992}
{Dupuis}, J., {Fontaine}, G., {Pelletier}, C., and {Wesemael}, F. (1992).
\newblock {A study of metal abundance patterns in cool white dwarfs. I -
  Time-dependent calculations of gravitational settling}.
\newblock {\em \apjs}, 82:505--521.

\bibitem[{Dupuis} et~al., 1993]{dupuis_fontaine_1993}
{Dupuis}, J., {Fontaine}, G., {Pelletier}, C., and {Wesemael}, F. (1993).
\newblock {A study of metal abundance patterns in cool white dwarfs. II -
  Simulations of accretion episodes}.
\newblock {\em \apjs}, 84:73--89.

\bibitem[{Farihi}, 2016]{farihi_2016}
{Farihi}, J. (2016).
\newblock {Circumstellar debris and pollution at white dwarf stars}.
\newblock {\em \nar}, 71:9--34.

\bibitem[{Farihi} et~al., 2008a]{farihi_2008a}
{Farihi}, J., {Becklin}, E.~E., and {Zuckerman}, B. (2008a).
\newblock {Spitzer IRAC Observations of White Dwarfs. II. Massive Planetary and
  Cold Brown Dwarf Companions to Young and Old Degenerates}.
\newblock {\em \apj}, 681:1470--1483.

\bibitem[{Farihi} et~al., 2011]{farihi_dufour_2011}
{Farihi}, J., {Dufour}, P., {Napiwotzki}, R., and {Koester}, D. (2011).
\newblock {The magnetic and metallic degenerate G77-50}.
\newblock {\em \mnras}, 413:2559--2569.

\bibitem[Farihi et~al., 2012]{farihi_gan2012}
Farihi, J., G{\"a}nsicke, B., Wyatt, M., Girven, J., Pringle, J., and King, A.
  (2012).
\newblock Scars of intense accretion episodes at metal-rich white dwarfs.
\newblock {\em Monthly Notices of the Royal Astronomical Society},
  424(1):464--471.

\bibitem[{Farihi} et~al., 2009]{farihi_jura_2009}
{Farihi}, J., {Jura}, M., and {Zuckerman}, B. (2009).
\newblock {Infrared Signatures of Disrupted Minor Planets at White Dwarfs}.
\newblock {\em \apj}, 694:805--819.

\bibitem[{Farihi} et~al., 2017]{farihi_vonHippel_2017}
{Farihi}, J., {von Hippel}, T., and {Pringle}, J.~E. (2017).
\newblock {Magnetospherically-trapped dust and a possible model for the unusual
  transits at WD 1145+017}.
\newblock {\em \mnras}, 471:L145--L149.

\bibitem[{Farihi} et~al., 2008b]{farihi_2008b}
{Farihi}, J., {Zuckerman}, B., and {Becklin}, E.~E. (2008b).
\newblock {Spitzer IRAC Observations of White Dwarfs. I. Warm Dust at
  Metal-Rich Degenerates}.
\newblock {\em \apj}, 674:431--446.

\bibitem[Ferrario et~al., 2015]{ferrario_demarco_2015}
Ferrario, L., de~Martino, D., and G{\"a}nsicke, B.~T. (2015).
\newblock Magnetic white dwarfs.
\newblock {\em Space Science Reviews}, 191(1):111--169.

\bibitem[{Ferrario} et~al., 2020]{ferrario_wickramasinghe_2020}
{Ferrario}, L., {Wickramasinghe}, D., and {Kawka}, A. (2020).
\newblock {Magnetic fields in isolated and interacting white dwarfs}.
\newblock {\em Advances in Space Research}, 66(5):1025--1056.

\bibitem[{Feuerbacher} and {Fitton}, 1972]{fuerbacher_fitton_1972}
{Feuerbacher}, B. and {Fitton}, B. (1972).
\newblock {Experimental Investigation of Photoemission from Satellite Surface
  Materials}.
\newblock {\em Journal of Applied Physics}, 43:1563--1572.

\bibitem[Fontaine and Brassard, 2008]{fontaine_and_brassard_2008}
Fontaine, G. and Brassard, P. (2008).
\newblock The pulsating white dwarf stars.
\newblock {\em Publications of the Astronomical Society of the Pacific},
  120(872):1043.

\bibitem[{Frewen} and {Hansen}, 2014]{frewen_hansen_2014}
{Frewen}, S.~F.~N. and {Hansen}, B.~M.~S. (2014).
\newblock {Eccentric planets and stellar evolution as a cause of polluted white
  dwarfs}.
\newblock {\em \mnras}, 439:2442--2458.

\bibitem[{Friedrich} et~al., 2000]{friedrich_koester_2000}
{Friedrich}, S., {Koester}, D., {Christlieb}, N., {Reimers}, D., and
  {Wisotzki}, L. (2000).
\newblock {Cool helium-rich white dwarfs from the Hamburg/ESO survey}.
\newblock {\em \aap}, 363:1040--1050.

\bibitem[G{\"a}nsicke et~al., 2012]{gansicke2012chemical}
G{\"a}nsicke, B., Koester, D., Farihi, J., Girven, J., Parsons, S., and Breedt,
  E. (2012).
\newblock The chemical diversity of exo-terrestrial planetary debris around
  white dwarfs.
\newblock {\em Monthly Notices of the Royal Astronomical Society},
  424(1):333--347.

\bibitem[{Garc{\'\i}a-Berro} et~al., 2016]{garcia-berro_kilic_2016}
{Garc{\'\i}a-Berro}, E., {Kilic}, M., and {Kepler}, S.~O. (2016).
\newblock {Magnetic white dwarfs: Observations, theory and future prospects}.
\newblock {\em International Journal of Modern Physics D}, 25:1630005.

\bibitem[Gentile~Fusillo et~al., 2018]{gentile2018can}
Gentile~Fusillo, N.~P., Tremblay, P.-E., Jordan, S., G{\"a}nsicke, B.~T.,
  Kalirai, J.~S., and Cummings, J. (2018).
\newblock Can magnetic fields suppress convection in the atmosphere of cool
  white dwarfs? a case study on wd2105- 820.
\newblock {\em Monthly Notices of the Royal Astronomical Society},
  473(3):3693--3699.

\bibitem[{Ghosh} and {Lamb}, 1979]{ghosh_lamb_1979}
{Ghosh}, P. and {Lamb}, F.~K. (1979).
\newblock {Accretion by rotating magnetic neutron stars. II. Radial and
  vertical structure of the transition zone in disk accretion.}
\newblock {\em \apj}, 232:259--276.

\bibitem[{Gregory} and {Donati}, 2011]{gregory_donati_2011}
{Gregory}, S.~G. and {Donati}, J.~F. (2011).
\newblock {Analytic and numerical models of the 3D multipolar magnetospheres of
  pre-main sequence stars}.
\newblock {\em Astronomische Nachrichten}, 332:1027.

\bibitem[{Gruen} et~al., 1994]{gruen_gustafson_1994}
{Gruen}, E., {Gustafson}, B., {Mann}, I., {Baguhl}, M., {Morfill}, G.~E.,
  {Staubach}, P., {Taylor}, A., and {Zook}, H.~A. (1994).
\newblock {Interstellar dust in the heliosphere}.
\newblock {\em \aap}, 286:915--924.

\bibitem[{Grun} et~al., 1984]{grun_morfill_1984}
{Grun}, E., {Morfill}, G.~E., and {Mendis}, D.~A. (1984).
\newblock {Dust-magnetosphere interactions}.
\newblock In {Greenberg}, R. and {Brahic}, A., editors, {\em IAU Colloq. 75:
  Planetary Rings}, pages 275--332.

\bibitem[{Hanu{\v{s}}} et~al., 2017]{hanus_viikinkoski_2017}
{Hanu{\v{s}}}, J., {Viikinkoski}, M., {Marchis}, F., {{\v{D}}urech}, J.,
  {Kaasalainen}, M., {Delbo'}, M., {Herald}, D., {Frappa}, E., {Hayamizu}, T.,
  {Kerr}, S., {Preston}, S., {Timerson}, B., {Dunham}, D., and {Talbot}, J.
  (2017).
\newblock {Volumes and bulk densities of forty asteroids from ADAM shape
  modeling}.
\newblock {\em \aap}, 601:A114.

\bibitem[Harrison et~al., 2018]{harrison_bonsor_2018}
Harrison, J. H.~D., Bonsor, A., and Madhusudhan, N. (2018).
\newblock {Polluted white dwarfs: constraints on the origin and geology of
  exoplanetary material}.
\newblock {\em Monthly Notices of the Royal Astronomical Society},
  479(3):3814--3841.

\bibitem[Hermes et~al., 2017]{hermes2017}
Hermes, J., G{\"a}nsicke, B., Kawaler, S.~D., Greiss, S., Tremblay, P.-E.,
  Fusillo, N.~G., Raddi, R., Fanale, S., Bell, K.~J., Dennihy, E., et~al.
  (2017).
\newblock White dwarf rotation as a function of mass and a dichotomy of mode
  line widths: Kepler observations of 27 pulsating da white dwarfs through k2
  campaign 8.
\newblock {\em The Astrophysical Journal Supplement Series}, 232(2):23.

\bibitem[{Hollands} et~al., 2015]{hollands_gainsicke_2015}
{Hollands}, M.~A., {G{\"a}nsicke}, B.~T., and {Koester}, D. (2015).
\newblock {The incidence of magnetic fields in cool DZ white dwarfs}.
\newblock {\em \mnras}, 450:681--690.

\bibitem[Hollands et~al., 2018]{hollands_gansicke_2018}
Hollands, M.~A., Gänsicke, B.~T., and Koester, D. (2018).
\newblock Cool dz white dwarfs ii: compositions and evolution of old remnant
  planetary systems.
\newblock {\em Monthly Notices of the Royal Astronomical Society},
  477(1):93--111.

\bibitem[{Hussain} and {Alecian}, 2014]{hussain_alecian_2014}
{Hussain}, G. A.~J. and {Alecian}, E. (2014).
\newblock {The role of magnetic fields in pre-main sequence stars}.
\newblock In {Petit}, P., {Jardine}, M., and {Spruit}, H.~C., editors, {\em
  Magnetic Fields throughout Stellar Evolution}, volume 302 of {\em IAU
  Symposium}, pages 25--37.

\bibitem[{Isakova} et~al., 2017]{isakova_zhilkin_2017}
{Isakova}, P.~B., {Zhilkin}, A.~G., {Bisikalo}, D.~V., {Semena}, A.~N., and
  {Revnivtsev}, M.~G. (2017).
\newblock {Features of the accretion in the EX Hydrae system: Results of
  numerical simulation}.
\newblock {\em Astronomy Reports}, 61:560--572.

\bibitem[{Joasil} et~al., 2017]{joasil_payne_2017}
{Joasil}, A., {Payne}, M.~J., and {Veras}, D. (2017).
\newblock {Planet-Planet Scattering and White Dwarf Pollution}.
\newblock In {\em American Astronomical Society Meeting Abstracts \#229},
  volume 229 of {\em American Astronomical Society Meeting Abstracts}, page
  433.20.

\bibitem[Jordan and Friedrich, 2002]{jordan2002}
Jordan, S. and Friedrich, S. (2002).
\newblock Search for variations in circular-polarization spectra of the
  magnetic white dwarf lp 790--29.
\newblock {\em Astronomy \& Astrophysics}, 383(2):519--523.

\bibitem[{Jura}, 2003]{jura_2003}
{Jura}, M. (2003).
\newblock {A Tidally Disrupted Asteroid around the White Dwarf G29-38}.
\newblock {\em \apjl}, 584:L91--L94.

\bibitem[{Jura}, 2008]{jura_2008}
{Jura}, M. (2008).
\newblock {Pollution of Single White Dwarfs by Accretion of Many Small
  Asteroids}.
\newblock {\em \aj}, 135:1785--1792.

\bibitem[{Jura} et~al., 2007]{jura_farihi_2007}
{Jura}, M., {Farihi}, J., and {Zuckerman}, B. (2007).
\newblock {Externally Polluted White Dwarfs with Dust Disks}.
\newblock {\em \apj}, 663:1285--1290.

\bibitem[{Kawka}, 2018]{kawka_2018}
{Kawka}, A. (2018).
\newblock {The properties and origin of magnetic fields in white dwarfs}.
\newblock {\em Contributions of the Astronomical Observatory Skalnate Pleso},
  48:228--235.

\bibitem[{Kawka} and {Vennes}, 2012]{KAWKA2012}
{Kawka}, A. and {Vennes}, S. (2012).
\newblock {VLT/X-shooter observations and the chemical composition of cool
  white dwarfs}.
\newblock {\em \aap}, 538:A13.

\bibitem[{Kawka} and {Vennes}, 2014]{kawka_vennes_2014}
{Kawka}, A. and {Vennes}, S. (2014).
\newblock {The polluted atmospheres of cool white dwarfs and the magnetic field
  connection}.
\newblock {\em \mnras}, 439:L90--L94.

\bibitem[{Kawka} et~al., 2019]{kawka_vennes_2019}
{Kawka}, A., {Vennes}, S., {Ferrario}, L., and {Paunzen}, E. (2019).
\newblock {Evidence of enhanced magnetism in cool, polluted white dwarfs}.
\newblock {\em \mnras}, 482(4):5201--5210.

\bibitem[{Kawka, A.} and {Vennes, S.}, 2011]{kawka_vennes_2011}
{Kawka, A.} and {Vennes, S.} (2011).
\newblock The cool magnetic daz white dwarf nltt~10480.
\newblock {\em A\&A}, 532:A7.

\bibitem[{Kepler} et~al., 2016]{sdss_dr12}
{Kepler}, S.~O., {Pelisoli}, I., {Koester}, D., {Ourique}, G., {Romero}, A.~D.,
  {Reindl}, N., {Kleinman}, S.~J., {Eisenstein}, D.~J., {Valois}, A.~D.~M., and
  {Amaral}, L.~A. (2016).
\newblock {New white dwarf and subdwarf stars in the Sloan Digital Sky Survey
  Data Release 12}.
\newblock {\em \mnras}, 455:3413--3423.

\bibitem[{Kepler} et~al., 2017]{kepler_romero_2017}
{Kepler}, S.~O., {Romero}, A.~D., {Pelisoli}, I., and {Ourique}, G. (2017).
\newblock {White Dwarf Stars}.
\newblock In {\em International Journal of Modern Physics Conference Series},
  volume~45 of {\em International Journal of Modern Physics Conference Series},
  page 1760023.

\bibitem[King, 1993]{king_1993}
King, A.~R. (1993).
\newblock {The accretion of diamagnetic blobs by a rotating magnetosphere}.
\newblock {\em Monthly Notices of the Royal Astronomical Society},
  261(1):144--148.

\bibitem[King and Regev, 1994]{king_regev_1994}
King, A.~R. and Regev, O. (1994).
\newblock {Spin rates and mass loss in accreting T Tauri stars}.
\newblock {\em Monthly Notices of the Royal Astronomical Society},
  268(1):L69--L73.

\bibitem[Kobayashi et~al., 2011]{kobayashi_2011}
Kobayashi, H., Kimura, H., Watanabe, S.-i., Yamamoto, T., and Müller, S.
  (2011).
\newblock Sublimation temperature of circumstellar dust particles and its
  importance for dust ring formation.
\newblock {\em Earth, Planets and Space}, 63(10):1067–1075.

\bibitem[{Koester}, 2009]{koester_2009}
{Koester}, D. (2009).
\newblock {Accretion and diffusion in white dwarfs. New diffusion timescales
  and applications to GD 362 and G 29-38}.
\newblock {\em \aap}, 498:517--525.

\bibitem[{Koester}, 2015]{koester_2015}
{Koester}, D. (2015).
\newblock {On Thermohaline Mixing in Accreting White Dwarfs}.
\newblock In {Dufour}, P., {Bergeron}, P., and {Fontaine}, G., editors, {\em
  19th European Workshop on White Dwarfs}, volume 493 of {\em Astronomical
  Society of the Pacific Conference Series}, page 129.

\bibitem[{Koester} and {Wilken}, 2006]{koester_wilken_2006}
{Koester}, D. and {Wilken}, D. (2006).
\newblock {The accretion-diffusion scenario for metals in cool white dwarfs}.
\newblock {\em \aap}, 453:1051--1057.

\bibitem[{Koester, D.} et~al., 2014]{koester_gansicke_2014}
{Koester, D.}, {G\"ansicke, B. T.}, and {Farihi, J.} (2014).
\newblock The frequency of planetary debris around young white dwarfs.
\newblock {\em A\&A}, 566:A34.

\bibitem[{Krzeminski} and {Serkowski}, 1977]{krzeminski_serkowski_1977}
{Krzeminski}, W. and {Serkowski}, K. (1977).
\newblock {Extremely high circular polarization of AN Ursae Majoris}.
\newblock {\em \apjl}, 216:L45--L48.

\bibitem[Kustler, 2007]{kustler2007}
Kustler, G. (2007).
\newblock Diamagnetic levitation-historical milestones.
\newblock {\em Revue Roumaine Des Sciences Techniques Serie Electrotechnique Et
  Energetique}, 52(3):265.

\bibitem[Lhotka and Galeş, 2019]{lhotka_2019}
Lhotka, C. and Galeş, C. (2019).
\newblock Charged dust close to outer mean-motion resonances in the
  heliosphere.
\newblock {\em Celestial Mechanics and Dynamical Astronomy}, 131(11).

\bibitem[{Li} and {Wilson}, 1999]{li_wilson_1999}
{Li}, J. and {Wilson}, G. (1999).
\newblock {Solutions to the Bernoulli Integral of the Funnel Flow}.
\newblock {\em \apj}, 527(2):910--917.

\bibitem[Malamud and Perets, 2019a]{malamud_perets_2019a}
Malamud, U. and Perets, H. (2019a).
\newblock Tidal disruption of planetary bodies by white dwarfs i: A hybrid
  sph-analytical approach.

\bibitem[Malamud and Perets, 2019b]{malamud_perets_2019b}
Malamud, U. and Perets, H. (2019b).
\newblock Tidal disruption of planetary bodies by white dwarfs ii: Debris disk
  structure and ejected interstellar asteroids.

\bibitem[Meintjes and Venter, 2005]{meintjes_2005}
Meintjes, P. and Venter, L. (2005).
\newblock The diamagnetic blob propeller in ae aquarii and non-thermal radio to
  mid-infrared emission.
\newblock {\em Monthly Notices of the Royal Astronomical Society},
  360(2):573--582.

\bibitem[Metzger et~al., 2012]{metzger_rafikov_2012}
Metzger, B.~D., Rafikov, R.~R., and Bochkarev, K.~V. (2012).
\newblock Global models of runaway accretion in white dwarf debris discs.
\newblock {\em Monthly Notices of the Royal Astronomical Society},
  423(1):505--528.

\bibitem[{Michalak}, 2000]{michalak_2000}
{Michalak}, G. (2000).
\newblock {Determination of asteroid masses --- I. (1) Ceres, (2) Pallas and
  (4) Vesta}.
\newblock {\em \aap}, 360:363--374.

\bibitem[{Mullally} et~al., 2007]{mullally_kilic_2007}
{Mullally}, F., {Kilic}, M., {Reach}, W.~T., {Kuchner}, M.~J., {von Hippel},
  T., {Burrows}, A., and {Winget}, D.~E. (2007).
\newblock {A Spitzer White Dwarf Infrared Survey}.
\newblock {\em \apjs}, 171:206--218.

\bibitem[{Mustill} et~al., 2018]{mustill_villaver_2018}
{Mustill}, A.~J., {Villaver}, E., {Veras}, D., {G{\"a}nsicke}, B.~T., and
  {Bonsor}, A. (2018).
\newblock {Unstable low-mass planetary systems as drivers of white dwarf
  pollution}.
\newblock {\em \mnras}, 476:3939--3955.

\bibitem[{Nauenberg}, 1972]{nauenberg_1972}
{Nauenberg}, M. (1972).
\newblock {Analytic Approximations to the Mass-Radius Relation and Energy of
  Zero-Temperature Stars}.
\newblock {\em \apj}, 175:417.

\bibitem[{Nixon} et~al., 2020]{nixon_pringle_2020}
{Nixon}, C.~J., {Pringle}, J.~E., {Coughlin}, E.~R., {Swan}, A., and {Farihi},
  J. (2020).
\newblock {Emission from elliptical streams of dusty debris around white
  dwarfs}.
\newblock {\em arXiv e-prints}, page arXiv:2006.07639.

\bibitem[{Nordhaus}, 2011]{nordhaus_2011}
{Nordhaus}, J. (2011).
\newblock {a Binary Scenario for the Formation of Strongly Magnetized White
  Dwarfs}.
\newblock {\em International Journal of Modern Physics E}, 20:29--36.

\bibitem[{Norton} et~al., 2008]{norton_butters_2008}
{Norton}, A.~J., {Butters}, O.~W., {Parker}, T.~L., and {Wynn}, G.~A. (2008).
\newblock {The Accretion Flows and Evolution of Magnetic Cataclysmic
  Variables}.
\newblock {\em \apj}, 672(1):524--530.

\bibitem[{Norton} et~al., 2004]{norton_wynn_2004}
{Norton}, A.~J., {Wynn}, G.~A., and {Somerscales}, R.~V. (2004).
\newblock {The Spin Periods and Magnetic Moments of White Dwarfs in Magnetic
  Cataclysmic Variables}.
\newblock {\em \apj}, 614(1):349--357.

\bibitem[{Rafikov}, 2011a]{rafikov_2011a}
{Rafikov}, R.~R. (2011a).
\newblock {Metal Accretion onto White Dwarfs Caused by Poynting-Robertson Drag
  on their Debris Disks}.
\newblock {\em \apjl}, 732:L3.

\bibitem[{Rafikov}, 2011b]{rafikov_2011b}
{Rafikov}, R.~R. (2011b).
\newblock {Runaway accretion of metals from compact discs of debris on to white
  dwarfs}.
\newblock {\em \mnras}, 416:L55--L59.

\bibitem[Reding et~al., 2020]{reding2020isolated}
Reding, J.~S., Hermes, J., Vanderbosch, Z., Dennihy, E., Kaiser, B., Mace, C.,
  Dunlap, B., and Clemens, J. (2020).
\newblock An isolated white dwarf with 317 s rotation and magnetic emission.
\newblock {\em The Astrophysical Journal}, 894(1):19.

\bibitem[Robertson, 1937]{robertson_1937}
Robertson, H. (1937).
\newblock Dynamical effects of radiation in the solar system.
\newblock {\em Monthly Notices of the Royal Astronomical Society}, 97:423.

\bibitem[{Schmidt} et~al., 1999]{schmidt_hoard_1999}
{Schmidt}, G.~D., {Hoard}, D.~W., {Szkody}, P., {Melia}, F., {Honeycutt},
  R.~K., and {Wagner}, R.~M. (1999).
\newblock {Accretion in the High-Field Magnetic Cataclysmic Variable AR Ursae
  Majoris}.
\newblock {\em \apj}, 525(1):407--419.

\bibitem[{Shahbaz}, 2019]{shahbaz_2019}
{Shahbaz}, T. (2019).
\newblock {Polarimetry of Binary Systems: Polars, Magnetic CVs, XRBs}.
\newblock In {Mignani}, R., {Shearer}, A., {S{\l}owikowska}, A., and {Zane},
  S., editors, {\em Astrophysics and Space Science Library}, volume 460 of {\em
  Astrophysics and Space Science Library}, page 247.

\bibitem[Stadel et~al., 2002]{stadel2002high}
Stadel, J., Wadsley, J., and Richardson, D.~C. (2002).
\newblock High performance computational astrophysics with pkdgrav/gasoline.
\newblock In {\em High Performance Computing Systems and Applications}, pages
  501--523. Springer.

\bibitem[Swan et~al., 2019]{swan_farihi_2019b}
Swan, A., Farihi, J., Koester, D., Hollands, M., Parsons, S., Cauley, P.~W.,
  Redfield, S., and Gänsicke, B.~T. (2019).
\newblock {Interpretation and diversity of exoplanetary material orbiting white
  dwarfs}.
\newblock {\em Monthly Notices of the Royal Astronomical Society},
  490(1):202--218.

\bibitem[{Swan} et~al., 2019]{swan_fraihi_2019}
{Swan}, A., {Farihi}, J., and {Wilson}, T.~G. (2019).
\newblock {Most white dwarfs with detectable dust discs show infrared
  variability}.
\newblock {\em \mnras}.

\bibitem[{Tout} et~al., 2008]{tout_wickramasinghe_2008}
{Tout}, C.~A., {Wickramasinghe}, D.~T., {Liebert}, J., {Ferrario}, L., and
  {Pringle}, J.~E. (2008).
\newblock {Binary star origin of high field magnetic white dwarfs}.
\newblock {\em \mnras}, 387:897--901.

\bibitem[Tremblay et~al., 2015]{tremblay2015evolution}
Tremblay, P.-E., Fontaine, G., Freytag, B., Steiner, O., Ludwig, H.-G.,
  Steffen, M., Wedemeyer, S., and Brassard, P. (2015).
\newblock On the evolution of magnetic white dwarfs.
\newblock {\em The Astrophysical Journal}, 812(1):19.

\bibitem[Ultchin et~al., 1997]{ultchin_regev_1997}
Ultchin, Y., Regev, O., and Bertout, C. (1997).
\newblock Diamagnetic blob interaction model of t tauri variability.
\newblock {\em The Astrophysical Journal}, 486(1):397--402.

\bibitem[Ultchin et~al., 2002]{ultchin_2002}
Ultchin, Y., Regev, O., and Wynn, G. (2002).
\newblock On the effects of the stellar magnetic field on the structure of t
  tauri accretion discs.
\newblock {\em Monthly Notices of the Royal Astronomical Society},
  331(3):578--586.

\bibitem[{Van Box Som} et~al., 2018]{vonboxsom_2018}
{Van Box Som}, L., {Falize}, {\'E}., {Bonnet-Bidaud}, J.~M., {Mouchet}, M.,
  {Busschaert}, C., and {Ciardi}, A. (2018).
\newblock {Numerical simulations of high-energy flows in accreting magnetic
  white dwarfs}.
\newblock {\em \mnras}, 473(3):3158--3168.

\bibitem[{Vanderbosch} et~al., 2020]{vanderbosch_hermes_2019}
{Vanderbosch}, Z., {Hermes}, J.~J., {Dennihy}, E., {Dunlap}, B.~H.,
  {Izquierdo}, P., {Tremblay}, P.~E., {Cho}, P.~B., {G{\"a}nsicke}, B.~T.,
  {Toloza}, O., {Bell}, K.~J., {Montgomery}, M.~H., and {Winget}, D.~E. (2020).
\newblock {A White Dwarf with Transiting Circumstellar Material Far outside the
  Roche Limit}.
\newblock {\em \apj}, 897(2):171.

\bibitem[{Veras} et~al., 2014]{veras_leinhardt_2014}
{Veras}, D., {Leinhardt}, Z.~M., {Bonsor}, A., and {G{\"a}nsicke}, B.~T.
  (2014).
\newblock {Formation of planetary debris discs around white dwarfs - I. Tidal
  disruption of an extremely eccentric asteroid}.
\newblock {\em \mnras}, 445:2244--2255.

\bibitem[{Veras} et~al., 2015]{veras_leinhardt_2015}
{Veras}, D., {Leinhardt}, Z.~M., {Eggl}, S., and {G{\"a}nsicke}, B.~T. (2015).
\newblock {Formation of planetary debris discs around white dwarfs - II.
  Shrinking extremely eccentric collisionless rings}.
\newblock {\em \mnras}, 451:3453--3459.

\bibitem[Veras and Wolszczan, 2019]{veras_wolszan_2019}
Veras, D. and Wolszczan, A. (2019).
\newblock {Survivability of radio-loud planetary cores orbiting white dwarfs}.
\newblock {\em Monthly Notices of the Royal Astronomical Society},
  488(1):153--163.

\bibitem[{Veras} et~al., 2018]{veras_xu_2018}
{Veras}, D., {Xu}, S., and {Rebassa-Mansergas}, A. (2018).
\newblock {The critical binary star separation for a planetary system origin of
  white dwarf pollution}.
\newblock {\em \mnras}, 473:2871--2880.

\bibitem[{Voss} et~al., 2007]{voss_koester_2007}
{Voss}, B., {Koester}, D., {Napiwotzki}, R., {Christlieb}, N., and {Reimers},
  D. (2007).
\newblock {High-resolution UVES/VLT spectra of white dwarfs observed for the
  ESO SN Ia progenitor survey. II. DB and DBA stars}.
\newblock {\em \aap}, 470:1079--1088.

\bibitem[{Wachlin, F. C.} et~al., 2017]{wachlin_vauclair_2017}
{Wachlin, F. C.}, {Vauclair, G.}, {Vauclair, S.}, and {Althaus, L. G.} (2017).
\newblock Importance of fingering convection for accreting white dwarfs in the
  framework of full evolutionary calculations: the case of the hydrogen-rich
  white dwarfs gd 133 and g 29-38.
\newblock {\em A\&A}, 601:A13.

\bibitem[{Wickramasinghe}, 2014]{wickramasinghe_2014}
{Wickramasinghe}, D. (2014).
\newblock {Accretion on to Magnetic White Dwarfs}.
\newblock In {\em European Physical Journal Web of Conferences}, volume~64,
  page 03001.

\bibitem[Wickramasinghe and Ferrario, 2000]{wickramasinghe_ferrario_2000}
Wickramasinghe, D. and Ferrario, L. (2000).
\newblock Magnetism in isolated and binary white dwarfs.
\newblock {\em Publications of the Astronomical Society of the Pacific},
  112(773):873--924.

\bibitem[{Wolff} et~al., 2002]{wolff_koester_2002}
{Wolff}, B., {Koester}, D., and {Liebert}, J. (2002).
\newblock {Element abundances in cool white dwarfs. II. Ultraviolet
  observations of DZ white dwarfs}.
\newblock {\em \aap}, 385:995--1007.

\bibitem[Wu, 2000]{Wu_2000}
Wu, K. (2000).
\newblock Accretion onto magnetic white dwarfs.
\newblock {\em Space Science Reviews}, 93(3):611--649.

\bibitem[{Wynn} and {King}, 1995]{wynn_king_1995}
{Wynn}, G.~A. and {King}, A.~R. (1995).
\newblock {Diamagnetic accretion in intermediate polars - I. Blob orbits and
  spin evolution}.
\newblock {\em \mnras}, 275:9--21.

\bibitem[Wynn et~al., 1997]{wynn_king_1997}
Wynn, G.~A., King, A.~R., and Horne, K. (1997).
\newblock {A magnetic propeller in the cataclysmic variable AE Aquarii}.
\newblock {\em Monthly Notices of the Royal Astronomical Society},
  286(2):436--446.

\bibitem[Xu and Jura, 2012]{xu_jura_2012}
Xu, S. and Jura, M. (2012).
\newblock {SPITZEROBSERVATIONS} {OF} {WHITE} {DWARFS}: {THE} {MISSING}
  {PLANETARY} {DEBRIS} {AROUND} {DZ} {STARS}.
\newblock {\em The Astrophysical Journal}, 745(1):88.

\bibitem[{Zhilkin} et~al., 2012]{zhilkin_bisikalo_2012}
{Zhilkin}, A.~G., {Bisikalo}, D.~V., and {Mason}, P.~A. (2012).
\newblock {Full 3D MHD calculations of accretion flow structure in magnetic
  cataclysmic variables with strong, complex magnetic fields}.
\newblock {\em Astronomy Reports}, 56:257--274.

\bibitem[{Zuckerman} et~al., 2011]{zuckerman_koester_2011}
{Zuckerman}, B., {Koester}, D., {Dufour}, P., {Melis}, C., {Klein}, B., and
  {Jura}, M. (2011).
\newblock {An Aluminum/Calcium-rich, Iron-poor, White Dwarf Star: Evidence for
  an Extrasolar Planetary Lithosphere?}
\newblock {\em \apj}, 739(2):101.

\bibitem[{Zuckerman} et~al., 2003]{zuckerman_koester_2003}
{Zuckerman}, B., {Koester}, D., {Reid}, I.~N., and {H{\"u}nsch}, M. (2003).
\newblock {Metal Lines in DA White Dwarfs}.
\newblock {\em \apj}, 596:477--495.

\bibitem[Zuckerman et~al., 2010]{zuckerman2010}
Zuckerman, B., Melis, C., Klein, B., Koester, D., and Jura, M. (2010).
\newblock Ancient planetary systems are orbiting a large fraction of white
  dwarf stars.
\newblock {\em The Astrophysical Journal}, 722(1):725.

\end{thebibliography}

\appendix

\section{Supplementary material \label{AP:SupMa}}
\begin{figure*}
\centering
\includegraphics[width=\textwidth]{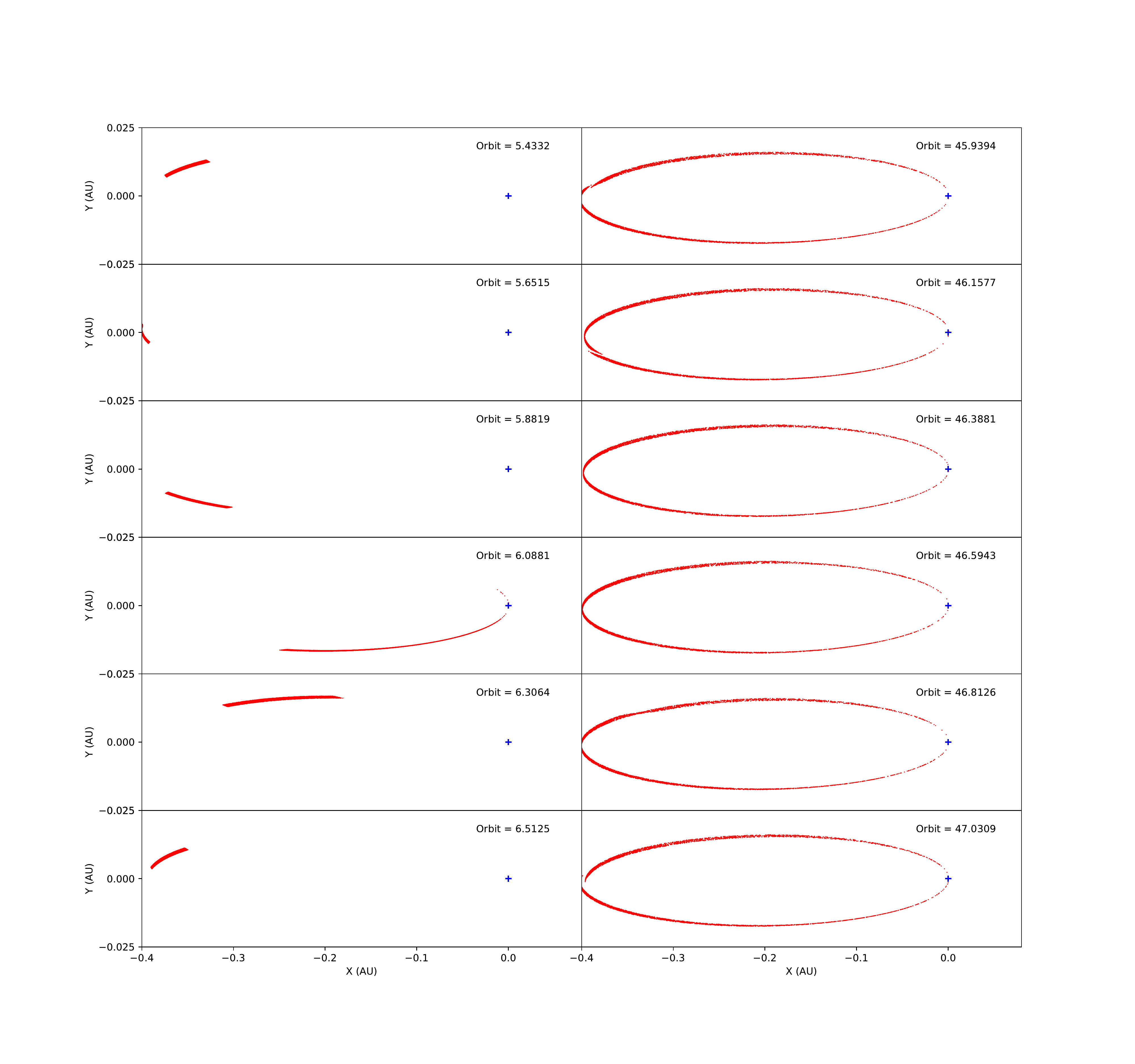}
\caption{The filling time of our disc after the tidal disruption using \texttt{Freefall}, showing that we are able to reproduce the results seen in figure 10 of \citet{veras_leinhardt_2014}.}
\label{fig:filling}
\end{figure*}

\begin{figure*}
    \centering
    \includegraphics[width=\textwidth]{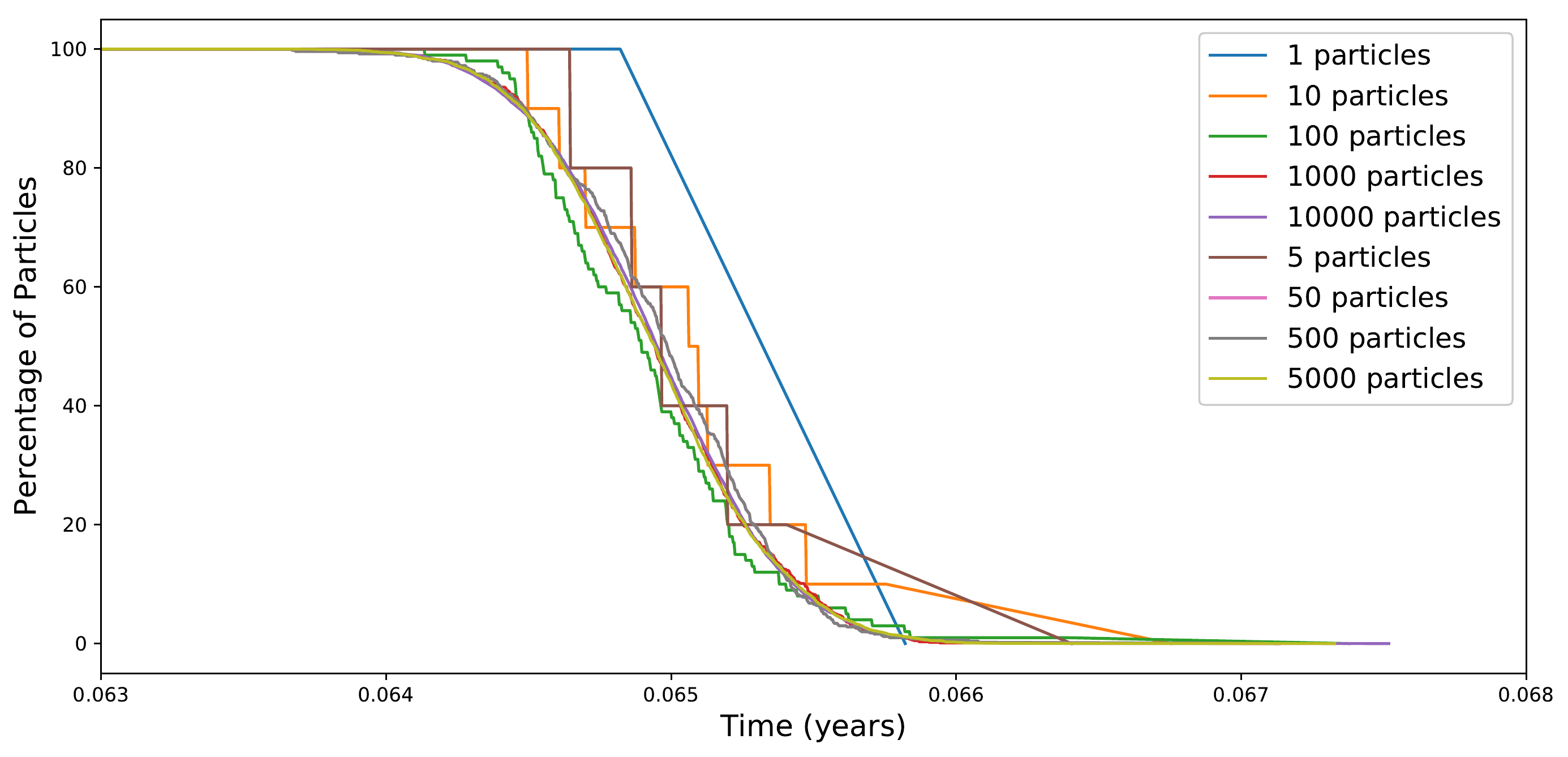}
    \caption{Lifetime of the disc for different initial particle number (50,100,500, 1000, and 5000). The graph shows the percentage of particles over time decreases identically in all three cases. Showing that we can use a smaller number of particles to speed up the simulation run in the lower magnetic field cases without affecting our results.}
    \label{fig:diff_part_num}
\end{figure*}{}

\begin{figure*}
    \centering
    \includegraphics[width=\textwidth]{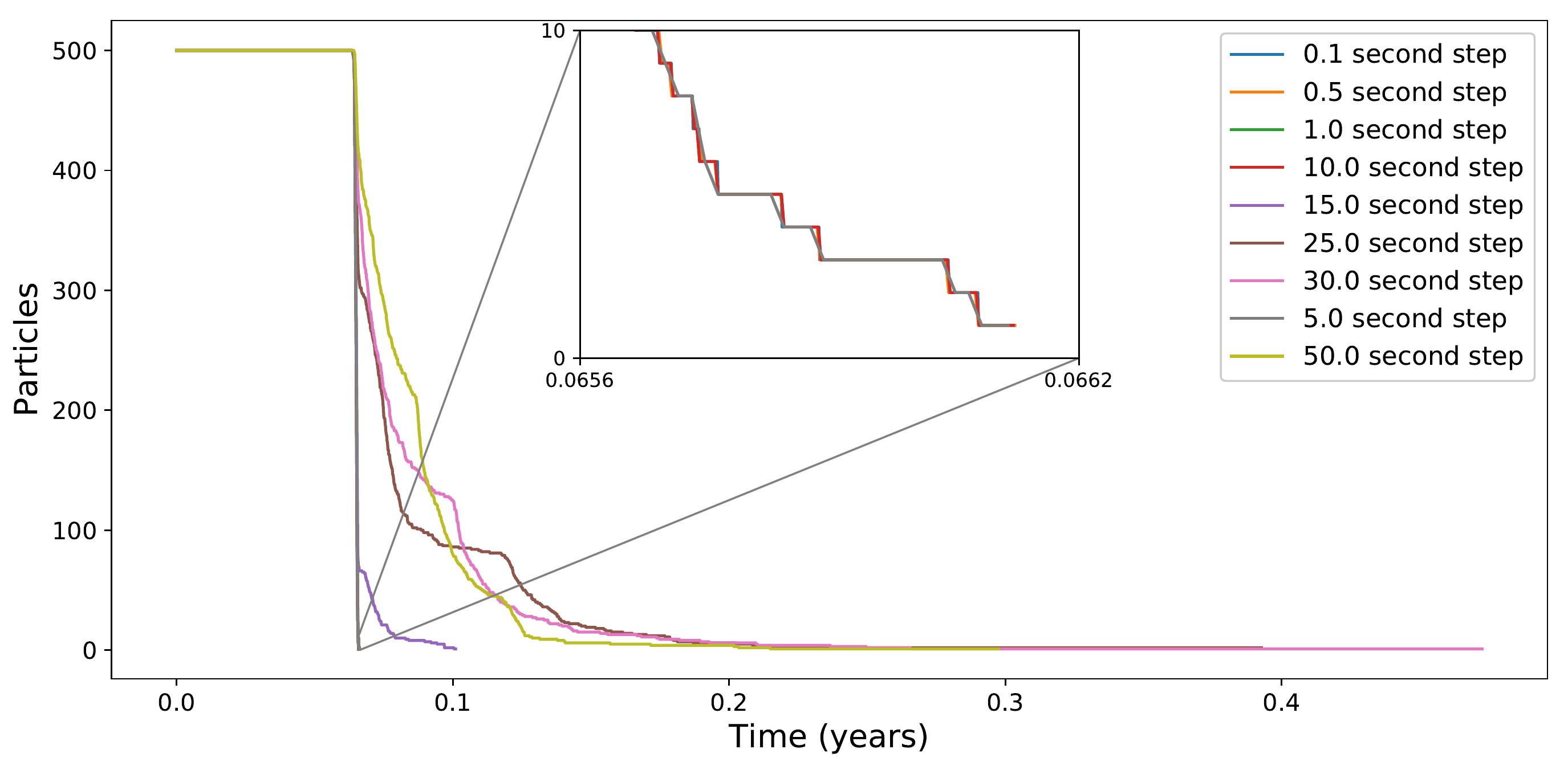}
    \caption{This figure shows the lifetime of 500 particles in the 1MG regime for different time steps. The results show that time steps below 10 seconds are all very similar. Above 10 seconds the results start to diverge and the particles are ejected instead of accreted. This shows that our simulation is numerically stable up to 10 seconds but cannot use time steps higher than this.}
    \label{fig:diff_time_step}
\end{figure*}{}

\end{document}